\documentclass[10pt]{iopart}

\expandafter\let\csname equation*\endcsname\relax
\expandafter\let\csname endequation*\endcsname\relax
\usepackage{amssymb}
\usepackage{amsmath}
\usepackage{natbib}

\usepackage{xcolor}
\usepackage{subfigure}
\usepackage{ctable}
\usepackage{makecell}
\usepackage{caption} 
\usepackage[shortlabels]{enumitem}
\usepackage{hyperref}
\usepackage{xspace}
\usepackage{gensymb}
\usepackage{lineno}

\usepackage{makecell}
\usepackage{multirow}
\usepackage{ctable}
\usepackage{ulem} 
\usepackage{bm}
\usepackage{algpseudocode}
\usepackage[ruled]{algorithm} 

\usepackage{xcolor}

\definecolor{darkpastelgreen}{rgb}{0.01, 0.75, 0.24}

\begin{document}

\title[DeepAstroUDA]{DeepAstroUDA: Semi-Supervised Universal Domain Adaptation for Cross-Survey Galaxy Morphology Classification and Anomaly Detection}

\author{A. \'Ciprijanovi\'c$^{1,2}$, A. Lewis$^{1}$, K. Pedro$^{1}$, S. Madireddy$^{3}$
B. Nord$^{1,2,4,5}$, G. N. Perdue$^{1}$, S. M. Wild$^{3,6}$} 

\address{${}^1$Fermi National Accelerator Laboratory, Batavia, IL 60510, USA}
\address{${}^2$Department of Astronomy and Astrophysics, University of Chicago, Chicago, IL 60637, USA}
\address{${}^3$Mathematics and Computer Science Division, Argonne National Laboratory, Lemont, IL 60439, USA}
\address{${}^4$Kavli Institute for Cosmological Physics, University of Chicago, Chicago, IL 60637, USA}
\address{${}^5$Laboratory for Nuclear Science, MIT, Cambridge MA, 02139, USA}
\address{${}^6$Applied Mathematics and Computational Research Division, Lawrence Berkeley National Laboratory, Berkeley, CA 94720, USA}

\ead{aleksand@fnal.gov}

\begin{abstract}
Artificial intelligence methods show great promise in increasing the quality and speed of work with large astronomical datasets, but the high complexity of these methods leads to the extraction of dataset-specific, non-robust features. Therefore, such methods do not generalize well across multiple datasets. We present a universal domain adaptation method, \textit{DeepAstroUDA}, as an approach to overcome this challenge.
This algorithm performs semi-supervised domain adaptation and can be applied to datasets with different data distributions and class overlaps. 
Non-overlapping classes can be present in any of the two datasets (the labeled source domain, or the unlabeled target domain), and the method can even be used in the presence of unknown classes.  
We apply our method to three examples of galaxy morphology classification tasks of different complexities ($3$-class and $10$-class problems), with anomaly detection: 1) datasets created after different numbers of observing years from a single survey (LSST mock data of $1$ and $10$ years of observations); 2) data from different surveys (SDSS and DECaLS); and 3) data from observing fields with different depths within one survey (wide field and Stripe 82 deep field of SDSS). For the first time, we demonstrate the successful use of domain adaptation between very discrepant observational datasets. \textit{DeepAstroUDA} is capable of bridging the gap between two astronomical surveys, increasing classification accuracy in both domains (up to $40\%$ on the unlabeled data), and making model performance consistent across datasets. Furthermore, our method also performs well as an anomaly detection algorithm and successfully clusters unknown class samples even in the unlabeled target dataset.

\end{abstract}


\ioptwocol
%

\section{Introduction}
\label{sec:intro}

With big datasets from current and next generation astronomical surveys like the Dark Energy Survey~\citep[DES;][]{DES2016}, the Hyper Suprime-Cam Subaru Strategic Program~\citep[HSC-SSP;][]{HSC2018}, the Vera Rubin Legacy Survey of Space and Time~\citep[LSST;][]{IK2019}, and the Nancy Grace Roman Space Telescope\footnote{\url{https://roman.gsfc.nasa.gov}}, development of artificial intelligence (AI) algorithms capable of combining knowledge from different telescopes will open doors to many new insights. 
For a review of the impact of AI on analysis of galaxy surveys, see~\cite{HCL2022}. 
Furthermore, many astrophysics and cosmology studies often start from simulated data that are well suited for learning the connection between the physical parameters and the observables. 
Simulations are the best (and sometimes the only) resource for creating labeled datasets needed for training supervised learning algorithms, which are ultimately intended to be used on real observational data. Unfortunately, standard deep learning algorithms are not well suited for working with multiple datasets. Very small, even pixel-level, differences between datasets can cause an algorithm trained on one dataset to experience a substantial drop in performance or even not work at all on another dataset~\citep{GD2016,DK2016,DK2017,FG2019,CK2021}. 
In astronomy, differences between datasets are often substantially larger. 
Simulated data can be different from observations due to computational constraints, approximations, missing or unknown physics, or imperfect addition of observational effects; while differences between observational datasets come from different telescope characteristics, observing times, and observing conditions.

Domain adaptation (DA) research includes the development of methods designed to bridge the gap between datasets and enable the creation of deep learning models that perform well on multiple datasets at the same time. This is done by guiding the model to learn and use only features present in both datasets~\citep{C2017,WD2018,GD2020}, that is, domain-invariant features. 
These methods can be divided into 1) distance-based methods such as Maximum Mean Discrepancy~\citep[MMD;][]{GB2007,GB2008}, Deep Correlation Alignment~\citep[CORAL;][]{SS2016}, Central Moment Discrepancy~\citep[CMD;][]{ZM2019}; and 2) adversarial-based methods such as Domain Adversarial Neural Networks~\citep[DANNs;][]{GU2016} and Conditional Domain Adversarial Networks~\citep[CDAN;][]{LC2017}. 

In astronomy, DA was first studied and used by~\cite{VD2019} for two different tasks: classification of Supernovae Ia and the identification of Mars landforms. 
Then, in our previous work~\cite{CK2021MNRAS}, DA was used for the problem of identifying merging galaxies in simulated and real data.
In this paper, we have used two DA techniques, MMD~\citep{GB2008} and DANNs~\citep{GU2016}, and have successfully trained a model that works on two Illustris-1~\citep{VG2014} simulated data sets of distant merging galaxies, as well as models that work on Illustris-1 simulated data of nearby merging galaxies and observed data from the Sloan Digital Sky Survey~\citep[SDSS;][]{YA2000}.
In~\cite{CK2021}, we showed that even smaller differences between datasets, such as inadvertent data perturbations that can naturally occur in astronomical data pipelines, can also cause the model trained on clean data to make catastrophic errors. 
Here, we developed deep learning models to classify galaxy morphology and include DA to increase model robustness and allow the model to perform well even in the presence of data perturbations. 

Furthermore,~\cite{GB2021} used an unsupervised instance-based domain adaptation method called the Kullback-Leibler Importance Estimation Procedure~\citep[KLIEP;][]{SN2007} to build a model that can more accurately derive star-formation histories from galaxy spectral energy distributions extracted from three different cosmological simulations: SIMBA~\citep{DA2019}, EAGLE~\citep{SC2015}, and IllustrisTNG~\citep{NS2019}. Finally, in~\cite{AG2021}, the authors use several DA methods to show that these techniques can significantly help with the drop-in performance when distinguishing between different types of dark matter substructures present in simulated strong gravitational lensing images of varying complexity.

Unfortunately, DA methods can be hard to fine tune and the choice of good hyperparameters can vary substantially for different datasets and types of studies.
They also often include the addition of multiple loss functions to the training procedure.
This leads to harder optimization tasks when searching for the true total loss function minima that will result in the best model performance. Finally, in real astrophysical applications, researchers will not always be able to work with nicely curated datasets. 
For example, when training models for classification tasks, we might encounter situations where two datasets contain only a portion of overlapping classes, while there might be other dataset-specific classes, not present in both datasets. 
Furthermore, datasets can also include unknown object classes, anomalies, or simply unlabeled portions of the data.
Most standard DA methods, such as MMD, try to align entire data distributions. 
The presence of any kind of non-overlapping classes makes the two data distributions very different, and so the DA method cannot be applied successfully. 

In this work, we develop a more universal DA method that does not require exact overlap between classes in the two datasets and can handle dataset-specific and non-overlapping classes in any of the two datasets. Furthermore, it works even in the presence of unknown classes and can even be used for anomaly detection. 
As with all DA methods, model training requires two datasets: the source data domain (with labeled images) and the target data domain (that can be unlabeled, as labels are not used during model training). 
Our aim is to develop a universal DA method that can be applied to a plethora of astronomical tasks and that can successfully perform DA on both simulated and astronomical survey data. 

Here we focus on galaxy morphology classification, generally into spiral, elliptical, and merging galaxies, and more granularly by leveraging sub-classes that more closely describe galaxy shapes (such as ellipticity, bulge prominence, presence of a bar in spiral galaxies, etc.). Understanding galaxy morphology is an important stepping stone for a full understanding of the formation and evolution of galaxies, mass assembly, and structure formation.
As galaxies evolve and interact with each other, complex structures are formed, such as bulges, spiral arms, bars, tidal tails, and clumps.
Morphology is also related to other physical properties of galaxies, such as their color, gas mass, stellar mass, and star formation rate~\citep{KH2003,LB2019}. 
The problem of galaxy morphology classification has been approached in several ways: visual inspection by experts~\citep{H1926,VB1960}; visual inspection by volunteers and crowd sourcing like in the Galaxy Zoo project~\citep{LS2008}; developing multiple sets of morphology parameters, such as the S\'ersic index~\citep{S1963}, or concentration, asymmetry, and clumpiness (CAS)~\citep{CB2003}; using simple machine learning algorithms on the morphology parameters~\citep{SR2019}; and even using more complex deep learning algorithms on the galaxy images themselves~\citep{WS2020,CB2021,CC2021}.
 
In this work, we develop a robust deep learning algorithm, capable of handling multi-dataset galaxy morphology problems. 
We used simulated LSST mock images made from IllustrisTNG~\citep{NS2019}, as well as multiple observational datasets available in the Galaxy Zoo project~\citep{LS2008,LS2011,WL2013}.
This way, we create a multi-dataset test bed well suited for the development of a universal DA method for galaxy morphology classification across different simulated and real datasets. 
Additionally, our method is equally applicable to other classification, regression, and anomaly detection tasks that use multiple datasets.

In Section~\ref{sec:methods}, we describe the DA method we develop and how it is implemented, and in Section~\ref{sec:network}, we describe the deep learning model we train to perform the galaxy morphology classification. In Section~\ref{sec:data}, we describe all of the datasets we use in our studies.
We present the results on different types of cross-dataset tasks in Section~\ref{sec:results}, with a discussion and conclusion in Section~\ref{sec:conclusion}. All the code used in this work can be found on our \href{https://github.com/deepskies/DeepAstroUDA}{GitHub page}\footnote{\url{https://github.com/deepskies/DeepAstroUDA}}, and we also make all of the datasets available on \href{https://doi.org/10.5281/zenodo.7473597}{Zenodo}\footnote{https://doi.org/10.5281/zenodo.7473597}.

\section{Methods}
\label{sec:methods}

DA methods are designed to help deep learning models extract and use only those features that are present in all datasets the model is intended to handle. Internally, using DA in model training helps align latent data distributions, allowing a model to find a common decision boundary between the classes for more consistent performance on multiple datasets~\citep{C2017,WD2018,GD2020}. Most distance-based DA methods (such as MMD or CORAL) and adversarial DA methods (such as DANNs) will only work efficiently under the assumption that the two latent data distributions are similar. These methods treat the entire data distribution as a whole; in other words, they are not class-aware. Therefore, correct alignment is only possible if both distributions contain the same classes of objects with similar properties. This condition is not always satisfied in scientific applications, so more flexible DA methods are needed.

\begin{figure}
    \centering
	\includegraphics[width=0.75\columnwidth]{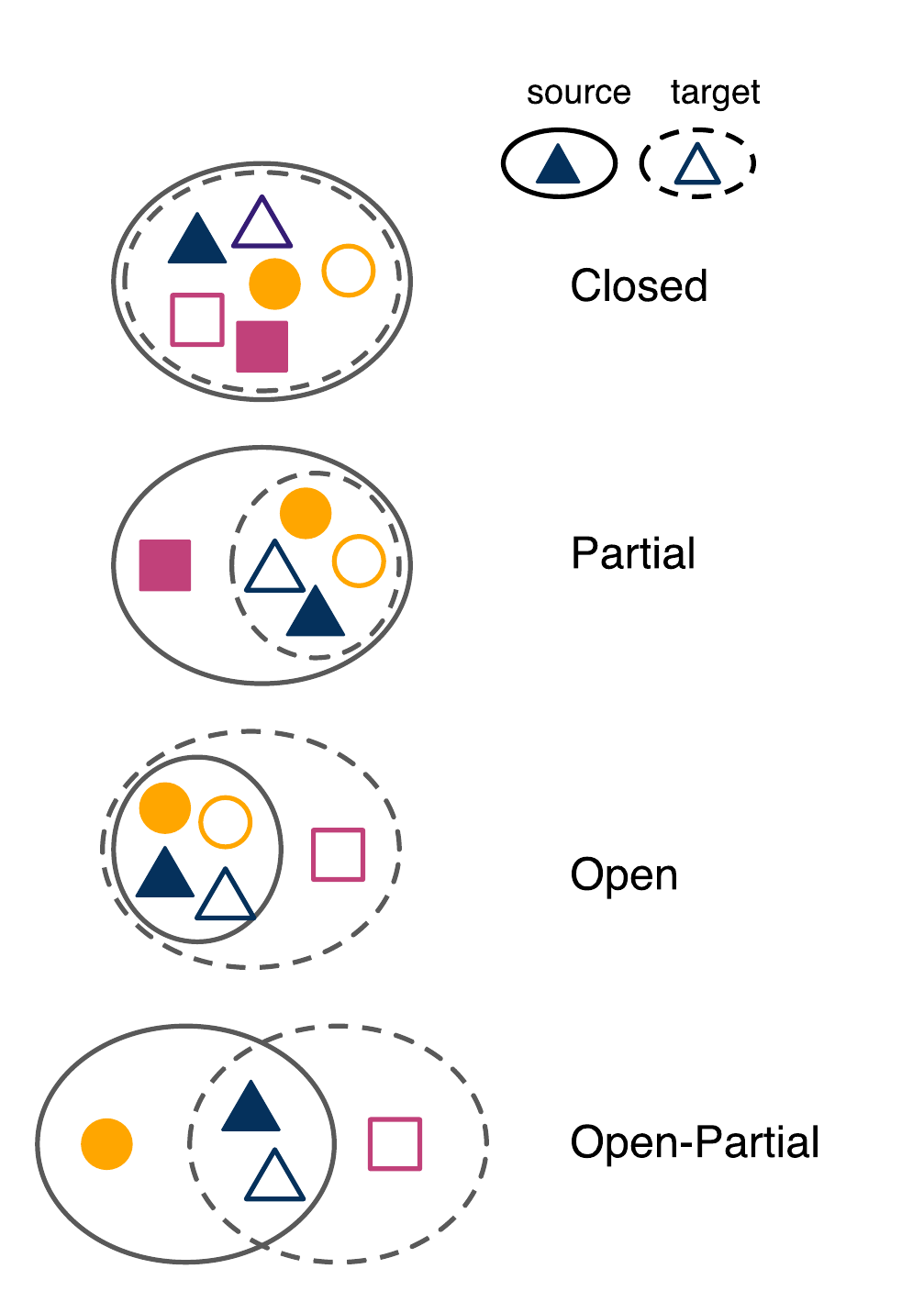}\\
    \caption{Types of DA problems. The source domain is represented with a solid line ellipse, while the target domain uses a dashed line ellipse. Source domain classes are represented by filled shapes and target domain by empty shapes.}
    \label{fig:da_types}
\end{figure}

We denote the source domain dataset as ${\cal D}_s=(X_s,Y_s)$ and the target domain dataset as ${\cal D}_t=(X_t,Y_t)$, where $X_s$ and $X_t$ are image sets from the source and target dataset, respectively, and $Y_s$ and $Y_t$ are the corresponding label sets. We can then divide class-aware DA methods into closed DA approaches that assume that the two datasets include the same classes, i.e. $Y_s = Y_t$~\citep{GB2008,GU2016}, and methods that assume that one of the domains contains more classes, such as open DA for $Y_s \subset Y_t$~\citep{BG2017,FW2019, LC2019,YH2022}, partial DA for $Y_t \subset  Y_s$~\citep{ZD2018,CY2019,XJ2021}, or DA for problems that are a mix of open and partial~\citep{YL2019}. For a graphical depiction of the different types of DA problems, see Figure~\ref{fig:da_types}. Since the inclusion of DA methods is needed when the target domain is unlabeled (otherwise one could simply train a regular supervised learning model on the target dataset directly), we may not know which of these situations will occur in our experiments. Hence, the development of more general methods, capable of handling all types of dataset overlaps (closed, open, partial, or mixed), is needed for full implementation of these methods in the sciences.

Here, we aim to develop such a Universal Domain Adaptation (UDA) method, capable of performing domain alignment even in the presence of extra classes (known or unknown). These additional classes can be present in both the labeled source or unlabeled target domain. The general principle is that extra classes, present in only one of the domains, should not be aligned with the other domain that does not contain them. This means that the unlabeled target domain can also contain completely unknown objects or anomalies that the method needs to recognize, group as a new class, and exclude from the domain alignment procedure. 

One such method, called Domain Adaptive Neighborhood Clustering via Entropy optimization (DANCE), was introduced in~\cite{SK2020}. Their method includes two novel ideas: 1) a neighborhood clustering technique to learn the structure of the target domain in a self-supervised way and to cluster neighboring source and target examples , and 2) using entropy to align the known or to reject unknown target classes. The authors show that DANCE can handle arbitrary domain shifts and they apply it to several benchmarking datasets, such as Office~\citep[$3$ domains and $31$ classes,][]{SK2010}, OfficeHome~\citep[$4$ domains and $65$ classes,][]{VE2017}, and VisDA~\citep[$2$ domains and $12$ classes,][]{PU2017}.

Furthermore, the idea of contrastive self-supervised learning has become a key component for learning semantically meaningful representations of the data in situation where no labels are available~\citep{WX2018,TK2019,SS2022,VS2022}. Representations are learned by comparing and contrasting positive (drawn by pairing the augmentations of the same image) and negative pairs of samples (different images) in an unsupervised setting without access to any labels. However, contrastive learning in the context of domain adaptation remains underexplored. In ~\cite{TL2021}, the authors propose to extend the contrastive learning approach to a situation where none of the data domains contain any labeled data. In our work, we combine ideas of contrastive learning on unlabeled samples and supervised learning using the labeled source domain data to build our UDA method for astronomy.

\subsection{DeepAstroUDA Method}
\label{sec:loss}
We introduce \textit{DeepAstroUDA}, which performs domain alignment and clustering of similar objects into classes via two loss functions: adaptive clustering and entropy separation. Clustering of target samples is performed using adaptive clustering ideas introduced in~\cite{LL2021}. To further improve the clustering of the unlabeled target domain, we use the entropy separation loss used in~\cite{SK2020}, but we improve upon this method by adding active hyperparameter tuning for this loss function~\footnote{We have also tested with the regular neighborhood clustering loss from DANCE, but we have found it is hard to fine tune and has worse performance then the adaptive clustering we describe below. We include both losses in the code available on our GitHub page, so an interested user can test both on their problem.}. 

\begin{figure}
    \centering
	\includegraphics[width=\linewidth]{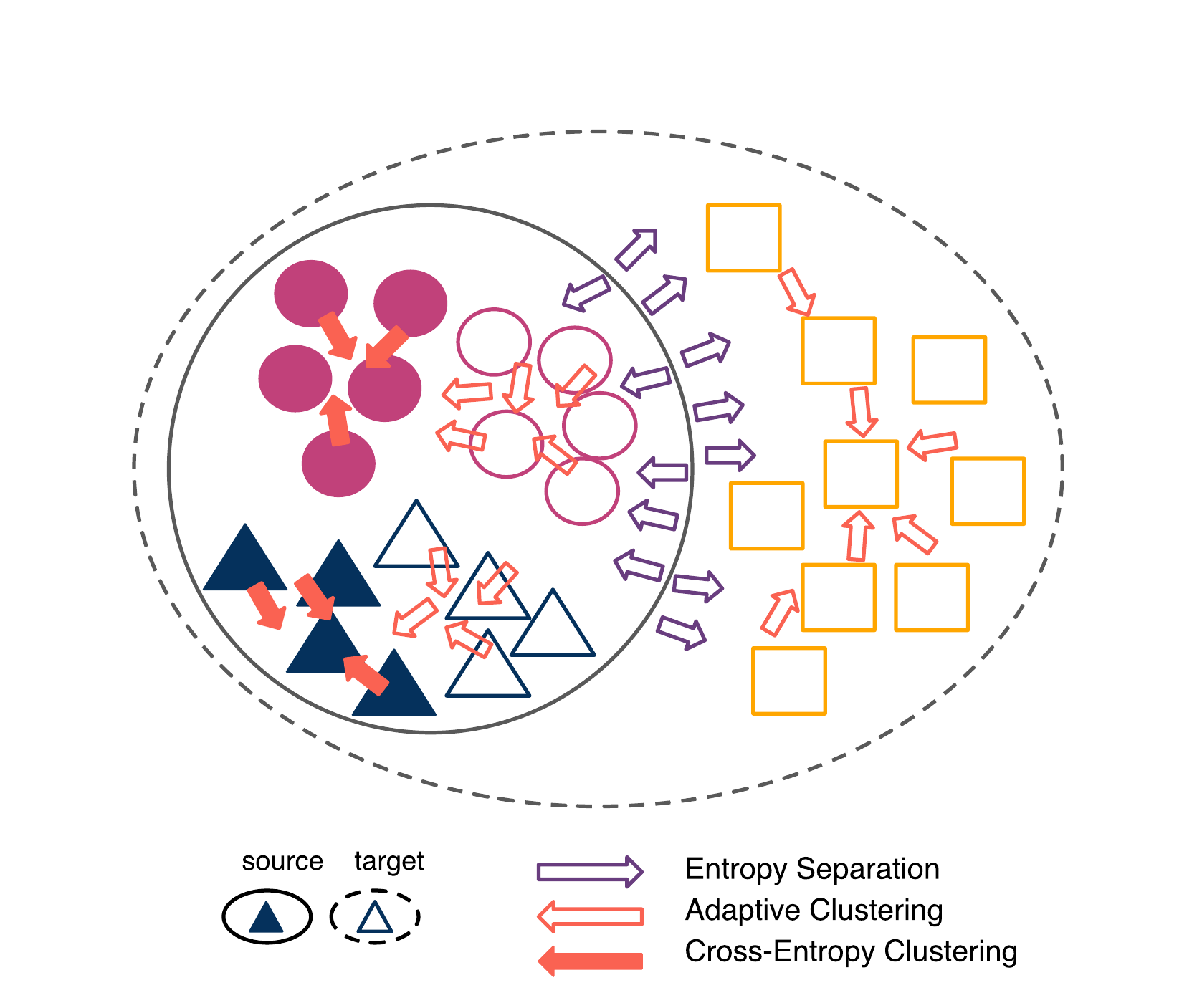}
    \caption{DeepAstroUDA method and the effects of different loss functions. The cross-entropy loss (filled red arrows) clusters the labeled source domain data (filled circles and triangles). The adaptive clustering loss (empty red arrows) pushes unlabeled target domain data (empty circles and triangles), towards data it shares the most similar features with (both source and target data). Finally, the entropy loss (empty violet arrows) uses entropy to push unknown classes away from the known ones. In this open DA example, the unknown class is present in the target domain (empty squares).}
    \label{fig:UDA}
\end{figure}

The power and flexibility of \textit{DeepAstroUDA}, compared to other methods from which we draw inspiration, come from both the combination of loss functions we use and from the active tuning of loss hyperparameters. Clustering of both known and unknown samples is performed in a self-supervised manner using contrastive learning ideas, which allows the model to compare all image pairs and cluster similar samples flexibly, without the need to align or understand the entire source and target data distributions. The entropy separation loss further enhances the rejection of very discrepant anomaly samples. Additional power and ease of use come from the active hyperparameter tuning, which maximizes performance of the DA loss as the training progresses and the source and target latent data distributions change. This circumvents the biggest challenge for most DA methods: finding well performing hyperparameters and maintaining good DA loss performance as data representations evolve.  In the following sections, we take a closer look at the different components of our \textit{DeepAstroUDA} method (see Figure~\ref{fig:UDA}) and how to fine-tune their hyperparameters for the best performance on multiple cross-dataset astrophysics applications. 

\subsection{Adaptive Clustering Loss}\label{sec:AC}

To build our Adaptive Clustering (AC) loss, we follow ideas from~\cite{LL2021} and ~\cite{SK2020}. The main idea of this type of semi-supervised clustering is to group target domain samples into clusters by computing pairwise similarities among features of unlabeled samples in the target domain. The loss minimization then forces the classifier to predict consistent class labels for samples with high pairwise feature similarities. This is achieved by training the model with a binary cross-entropy loss, where binary pairwise feature similarities are used as ground truth labels.

To facilitate semi-supervised clustering and to increase the number of similar samples we perform a data augmentation step. For each image in the source and target batches, we create two transformed versions ($90^{\circ}$ rotation and scaling by zooming to 300 pixels than cropping back to 256) of that image. Then, for any pair of unlabeled target samples $x_1$ and $x_2$, we predict a pairwise similarity label by using the classifier output prediction vectors $\bm p_1$ and $\bm p_2$ and rank ordering their elements~\citep{HR2020}. We then require the top $k$ elements to have the same ordering ($k=3$ for 3-class problem and $k=7$ for 10-class problem, for computational efficiency) to decide that the paired samples belong to the same class, which is denoted by a similarity label $s_{12} = 1$; otherwise, $s_{12} = 0$. For the labeled source domain images, we use their class labels to generate similarity labels. We also calculate a similarity score between samples as $\hat{s}_{12}=\bm p^\top_1 \bm p_2$. Finally, we can write the AC loss as a binary cross-entropy loss, where similarity labels are used as ground truth labels:
\begin{equation}
{\cal L}_\mathrm{AC} = -\sum_{i \in B}\sum_{j \in b_t} s_{ij}\mathrm{log}(\hat{s}_{ij}) + (1-s_{ij})\mathrm{log}(1-\hat{s}_{ij}),
\end{equation}
where $B$ is the bank that contains samples from all previous source and target batches, and $b_t$ is the current target batch~\citep{SK2020}. By comparing similarities between unlabeled target samples from the current target batch to all elements stored in the bank, current target samples are pushed towards the source and target samples with which they share the most similar features.

\subsection{Entropy Separation Loss}\label{sec:ES}

Entropy Separation (ES) has an explicit objective to encourage the alignment (known classes, present in both domains) or rejection (unknown classes present only in one of the domains) of target samples~\citep{SK2020}. This is possible because unknown samples often do not share features with known samples, which leads to larger entropy of the classifier output for unknown samples compared to entropy of the output for known classes~\citep{YL2019}. Therefore, the entropy can be used to decide the boundary between known and unknown samples. If we denote the mean entropy of the classifier output $\bm p_i$ of sample $i$ from the target batch $b_t$ as $H(\bm p_i)$, we can define a boundary $\rho$ around the entropy value so that:

\begin{equation}
{\cal L}_\mathrm{ES}(\bm p_i) = 
\begin{cases}
-|H(\bm p_i)-\rho| &  |H(\bm p_i)-\rho| > m,\\
0 & \mathrm{otherwise}.
\end{cases}
\end{equation}

Here $m$ is a confidence threshold around the boundary $\rho$, which is used to decide if we are confident about whether a particular sample belongs in a known or unknown class. Only those samples that are far enough from the entropy boundary $\rho$ will be moved towards known classes or pushed away as an unknown class. The boundary $\rho$ and confidence threshold $m$ start from preset values (determined from experiments), but are actively fine-tuned during training. Section~\ref{sec:tuner} describes the active fine-tuning of parameters in more detail. Finally, the total ES loss is:
\begin{equation}
{\cal L}_\mathrm{ES} = \frac{1}{|b_t|}\sum_{i \in b_t} {\cal L}_\mathrm{ES}(\bm p_i). 
\end{equation}

\subsection{Total Loss and Model Training}

\begin{figure}
\begin{centering}
	\includegraphics[width=\linewidth]{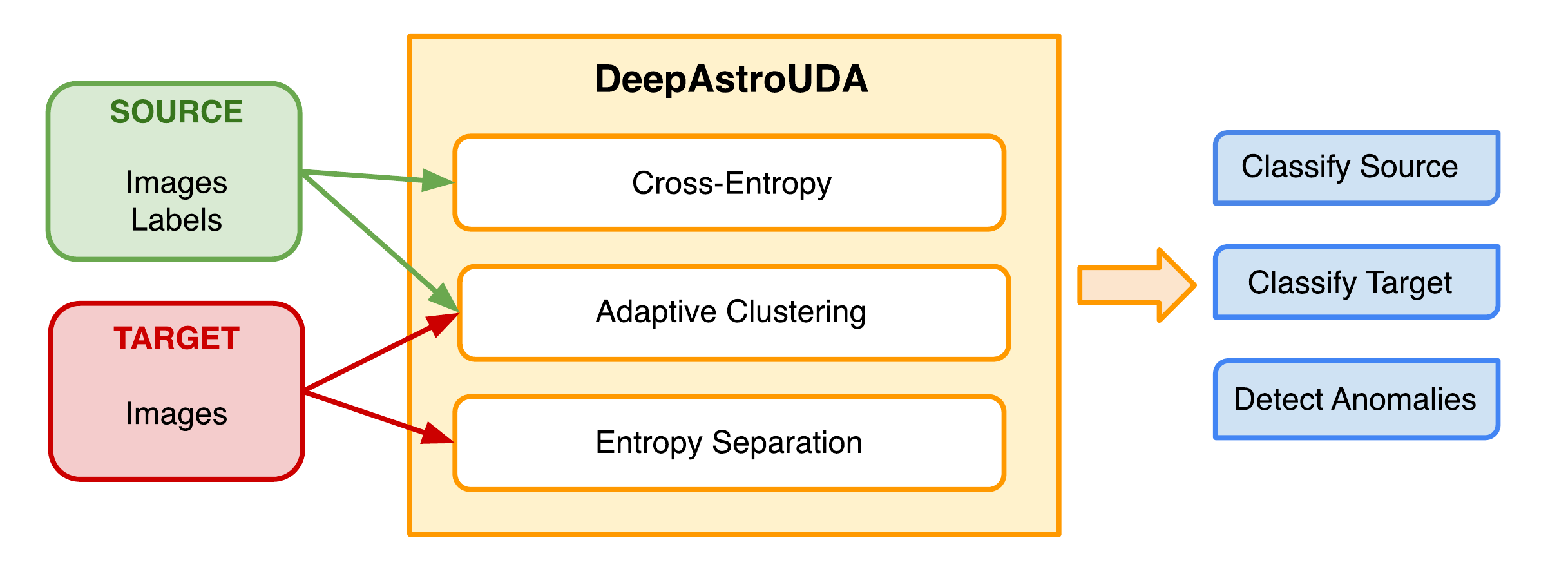}
    \caption{A high-level schematic of the \textit{DeepAstroUDA} architecture.}
    \label{fig:framework}
    \end{centering}
\end{figure}

While the ES loss is applied only to the target domain data, the AC loss is used to cluster the target data by aligning the samples to both the source and target data via the bank that stores all previously seen samples from both source and target batches. 
Finally, the main classification loss, applied only to the labeled source domain data, is the standard weighted Cross-Entropy (CE) loss:
\begin{equation}
{\cal L}_\mathrm{CE}= \frac{- \sum\limits_{k=1}^{\mathrm{K}} w_k y_k \log \hat{y}_k}{\sum\limits_{k=1}^{\mathrm{K}} w_k},
\end{equation}
where the class weight (distinct from the network weight parameters) for each class is calculated as $w_k = \mathrm{N_s}/ (\mathrm{K} n_k)$, where $n_k$ is the number of images in class $k$, $\mathrm{K}$ is the total number of source classes, and $N_s$ is the total number of images in the training source dataset. The true and predicted labels are $y_k$ and $\hat{y}_k$, respectively. See Figure~\ref{fig:framework} for a diagram showing how different loss functions within \textit{DeepAstroUDA} utilize the source and target datasets.

The final objective of the model training is: 
\begin{equation}
{\cal L} = {\cal L}_{CE} + \lambda({\cal L}_\mathrm{AC} + {\cal L}_\mathrm{ES}).
\end{equation}

We utilize domain-specific batch normalization, which eliminates domain style information leakage, which can be viewed as a form of weak domain alignment. Finally, the importance of the clustering losses that perform domain adaptation is governed by the DA weight parameter $\lambda$. We tested values in the range of $0.0001-100$ and find that $\lambda =0.005$ is a good value that achieves the best model performance. Example of the behaviour of each of the three losses during training is shown in Figure~\ref{fig:losses} in Section~\ref{sec:within_survey_LSST_Y1Y10}

\subsection{Hyperparameter Tuner}\label{sec:tuner}

Fine-tuning network hyperparameters can be a challenging task. Most DA methods include multiple hyperparameters inside different loss functions, as well as loss weights that control the contribution of each loss term to the total loss function used in the training. Often, the final model performance is very dependent on the hyperparameters. In most cases, the hyperparameter values are hardcoded for a specific dataset and use case scenario, which is not useful when applying the same algorithm to a different dataset. Furthermore, to determine the optimal hyperparameter values, we often rely on simple random or grid search, or Bayesian optimization~\citep{BB2011,SS2016bayes} over the parameter space, which can be slow or require strong computational resources. More importantly, as DA training progresses and the source and target latent data representations change, the choice of optimal hyperparameters might also change, which is not taken into account in most DA approaches.

In this work we develop and utilize a hyperparameter tuner that actively changes and fine-tunes the hyperparameters of the ES loss to help successfully cluster unlabeled target samples (see Algorithm~\ref{algo:tuner}). It focuses on two parameters: the boundary $\rho$ around the sample entropy value and the confidence interval $m$ (see Section~\ref{sec:ES}). During training, our hyperparameter tuner iteratively checks the performance of the model and changes hyperparameter values to improve it. It first changes the boundary $\rho$ and then adjusts the confidence $m$. Once the loss stops improving, the tuner repeats this cycle. We found that active hyperparameter tuning of the ES loss is crucial for good performance, since both the boundary and confidence interval values will certainly change as the training progresses, the target samples become more clustered, and the distinction between classes becomes clearer.

The initial guess for the boundary $\rho$ is calculated as $\rho=\log(K)/2$, where $K$ is the number of known source classes. The initial value for $m$ and the step value used by the tuner for both $\rho$ and $m$ are found through experimentation. We run experiments for problems between $3$ and $10$ classes to set the initial parameter values for the tuner. For example, we found that for problems with less than $5$ classes initial values for $m$ are usually in the range of $0.2-0.8$, while for more classes the range is $1.1-1.6$. Once the training is performed on any new dataset, the tuner will use the closest values for the specific class-size problem. Since the preset initial value is not necessarily the best choice for all datasets, using active tuning is encouraged. Different initial tuner values and step sizes for both parameters to further improve the performance can be specified manually, and other hyperparameters can be added to the tuner.

\begin{algorithm}
  \caption{Hyperparameter tuner.}\label{algo:tuner}
  \textbf{Input} number of known classes $K$ \\
  \textbf{Output} Entropy loss parameters $\rho$ and $m$ 
  \begin{algorithmic}
      \State $\rho\gets \log(K)/2$
      \State $m\gets m_0$ \Comment{Initial $m_0$ found experimentally}
      \State $j\gets 0$ \Comment{Step counter}
      \State $step\gets [0.3,-0.3,0.5,-0.5]$ \Comment{Step sizes}
      \State $epoch\gets 0$  \Comment{Epoch counter}
      \State $change\gets 0$  \Comment{Change counter}\\
      \(\triangleright\) After each change, epoch counter is reset and change counter is incremented
      \While {training}
        \State $epoch{+}{+}$
        \If{${\cal L}>= {\cal L}_\mathrm{min}$}
            \If{$change==0$ and $epoch>5$}
                \State $\rho\gets \rho+step[j]$
            \ElsIf{$change==1$ and $epoch>2$} 
                \State $m\gets m+step[j]$
            \ElsIf{$change==2$ and $epoch>2$}
                \State $m\gets 2m$
            \ElsIf{$change==3$ and $epoch>2$}
                \State $j\gets j+1$
                \State $change \gets 0$
            \EndIf
        \EndIf
      \EndWhile
  \end{algorithmic}
\end{algorithm}

\subsection{Latent Space Visualization}

Visualization of the latent data distribution is important to understand the model behavior, performance, and trustworthiness, and ultimately to conduct further model refinement. It is particularly useful in DA tasks, where the model is trained to align data distributions from two different data domains. Furthermore, our method not only tries to align classes present in both data domains, but also pushes away unknown samples or classes that are present in only one of the domains and therefore should not be used in the domain alignment. This makes latent space visualizations even more important.

To visualize the latent data distributions, we utilize manifold learning methods because of their capability to recognize non-linear data patterns. We use a combination of the isomap~\citep{TS2000} and the t-distributed stochastic neighbor embedding~\citep[tSNE;][]{MH2008}. 

An isomap is a lower-dimensional embedding of the latent space, such that geodesic distances in the original higher-dimensional space are also respected in the lower-dimensional space. It is constructed in three stages. First, a weighted neighborhood graph $G$ over all data points is constructed. Then, the edge weight values of the graph $G$ are assigned the distances between neighboring points and the geodesic distances between all pairs of points are estimated as their shortest-path distances in the graph $G$. Finally, the lower $d$-dimensional (often $d=2$ or $d=3$ for visualization purposes) embedding that best preserves the manifold’s estimated intrinsic geometry is produced by applying classical Metric Multidimensional Scaling ~\citep[MDS;][]{BG2005} to the matrix of the shortest-graph distances. We use the \texttt{scikit} implementation of isomaps~\citep{PV2011}. Isomaps are a great tool for realistic latent data representation, but they do not always show the cleanest and most visually distinct clumps for different classes, which can make interpretation harder.

The t-SNE method calculates the probability distribution over data point pairs in the latent space of the model. It assigns a higher probability to similar objects and a lower probability to dissimilar pairs. It then constructs a 2D or 3D representation of the data, in which data pairs share the same probabilities. By minimizing the Kullback–Leibler (KL) divergence~\citep{KL1951} between the two distributions, the t-SNE method ensures the similarity between the actual distribution and the low-dimensional projection. The t-SNE algorithm on its own can be slow, but it produces well separated clumps for different classes. Furthermore, since it adapts to data by performing different transformations in different regions, it is difficult to compare the relative sizes of clusters in t-SNE plots. In addition, the final appearance of the t-SNE plots is highly dependent on several user-defined parameters~\cite{WV2016}. 

In this work, we combine isomaps and t-SNE plots and utilize the best properties of both methods. First, we create a 2D isomap representation of the latent space of the model, ensuring that the intrinsic geometry of the original high-dimensional space is preserved. We then use the isomap as an input to the t-SNE algorithm, to create a more visually distinct classes that are easier to interpret. Since t-SNE in this case does not need to further reduce the dimensionality of the space, we do not expect dramatic changes in the positions of the class clumps. In the following text, we refer to our visualizations as t-SNE plots, but we emphasize here that the presented plots actually show t-SNE plots of the isomaps of the latent space of the model.

\section{Neural Network Model and Training}
\label{sec:network}
In all of our experiments, we use the \textit{ResNet50}~\citep{HZ2016} network (with random initialization of model weights). We train it with early stopping, which monitors the classification accuracy and stops the training when there is no improvement after 12 consecutive epochs. The model is trained using stochastic gradient descent with Nesterov momentum~\citep{SM2013} and an initial learning rate of $0.001$. The learning rate is tuned using an inverse learning rate scheduler, whereby the learning rate is decayed by a factor of $0.1$ every set number of epochs, determined by the specific initial dataset configuration (every $10$ epochs in all of our 10-class experiments and every $7$ epochs for our 3-class experiment). We train our models on 4 NVIDIA RTX A6000 GPUs (available from Google Colab and LambdaLabs), and on average, the training converges in ${\approx}5$ hours (dependent on the dataset size and the complexity of the experiment).

\section{Data}
\label{sec:data}

\subsection{Simulations - LSST Mock Data}
\label{sec:data_simulated}

IllustrisTNG~\citep{MV2018,NP2018,SP2018,NP2018,PH2018,NS2019} is a state-of-the-art cosmological magneto-hydrodynamical simulation that includes gas, stars, dark matter, supermassive black holes, and magnetic fields. It uses a galaxy formation model built on the cosmological simulation code AREPO~\citep{S2010}, which solves the coupled equations of ideal magneto-hydrodynamics and self-gravity. IllustrisTNG builds on the successes of the older simulation, Illustris-1~\citep{VG2014}, but extends the mass range of the simulated galaxies and halos, uses improved numerical and astrophysical modeling, and addresses some of the identified shortcomings and tensions with observations (see~\cite{NP2015}).

The Illustris project is ideally suited for studying galaxy formation and morphology, merging galaxies, gas accretion, formation of galaxy clusters, and large-scale structures. It is also a great resource for the creation of labeled mock datasets that can be made to mimic different astronomical surveys and used to train different kinds of supervised learning algorithms.

We use simulated LSST mocks from~\cite{CK2021}, which are made from the IllustrisTNG100~\citep{NS2019} simulation. Mock images include three filters ($g,r,i$), made from the two simulation snapshots: 95 (redshift $z=0.05$) and 99 (redshift $z=0$). All images were converted to an effective redshift of $z=0.05$, to create a larger single-redshift dataset. This dataset includes three galaxy morphology classes---spiral, elliptical and merging galaxies---made by following~\cite{LP2004} and~\cite{ST2015}, who use the $G$-$M_{20}$ ``bulge statistic'' to distinguish between galaxy morphology classes. It also includes augmentation of the least numerous class (merging galaxies) to create the final balanced dataset of ${\approx}35,000$ galaxy images (with dimensions $100\times100$ pixels), which is the approach we follow in this work.

The creation of LSST mock observations from IllustrisTNG100 was done using the \texttt{GalSim} package~\citep{RJ2015}. We use both datasets created in~\cite{CK2021}---a high-noise one-year survey (``Y1'') and a low-noise ten-year survey (``Y10'')---made by applying an exposure time corresponding to one year or ten years of observations directly to the raw images ($552\,\mathrm{s}$ per year for the $r$ and $i$ filters and $240\,\mathrm{s}$ for the $g$ filter). The images also include both atmospheric and optical point spread function (PSF) blurring, mimicking LSST observations. Finally, the images include arcsinh stretching to make fainter objects more apparent while preserving the original color ratios in each pixel. For more details on the mock dataset creation, see~\cite{CK2021}.

\subsection{Observations}
\label{sec:data_observed}

The Galaxy Zoo project~\citep[GZ;][]{LS2008} was the first to provide morphological classifications of nearly one million galaxies from the Sloan Digital Sky Survey~\citep[SDSS;][]{YA2000}, performed by ${\approx}10^5$ volunteers that classified images using a web-based interface. While the first project used a simplified classification into spiral, elliptical and merging galaxies, its successor, Galaxy Zoo 2~\citep[GZ2;][]{WL2013}, used a more complex classification system that considered the presence of bars and bulges, the shapes of edge-on disks, and quantification of the relative strengths of galactic bulges and spiral arms. It provided more that $300,000$ reliable morphological classifications of galaxies in the full Data Release 7 (DR7) of SDSS as well as the deeper Stripe 82. Following the success of these volunteer-based projects, the Galaxy Zoo expanded to include several other classification-based projects that use data from other telescopes and astronomical surveys\footnote{For decision trees used in different GZ projects, see \url{https://data.galaxyzoo.org/gz_trees/gz_trees.html}}. The newer projects include Galaxy Zoo 3: Hubble~\citep{WG2017}; Galaxy Zoo 4: CANDELS~\citep{SL2017}, DECaLS~\citep{WL2022}, UKIDSS~\citep{G2017}, Ferengi~\citep{G2017}, GAMA-KiDS~\citep{HB2019A}; and Galaxy Zoo Illustris~\citep{DF2018}, using simulated data.

In this work, we apply our model to several GZ datasets to test its performance on different observational cross-dataset scenarios and use cases.

\subsubsection{Galaxy Zoo: SDSS and DECaLS\\}
\label{sec:GZ2_SDSS_DECaLS}
We use two GZ datasets~\citep{LS2008,WL2013}\footnote{Current publicly available GZ datasets can be found at \url{https://data.galaxyzoo.org}}: the source domain dataset is from GZ2 SDSS~\citep{LS2008,LS2010,WL2013}, and the target domain dataset is from GZ3 DECaLS~\citep{WL2022}, which uses DR7 of the the DECam Legacy Survey ~\cite[part of the Dark Energy Spectroscopic Instrument (DESI) Legacy Imaging Surveys,][]{DS2019}.

Specifically, for the target domain, we use an ${\approx}18$k subset of the DECaLS dataset, with images that passed more rigorous vote filtering and better class separation, named Galaxy10 DECaLS\footnote{Galaxy10 data is available at \url{https://astronn.readthedocs.io/en/latest/galaxy10.html}}. We perform a 10-class experiment, in which we use 9 classes from this dataset (disturbed, merging, round smooth, cigar-shaped smooth, barred spiral, unbarred tight spiral, unbarred loose spiral, edge-on without bulge, edge-on with bulge). We add one more class from the full GZ3 DECaLS dataset, gravitationally lensed galaxies, which we will treat as an unknown class present in the target domain. We chose gravitational lenses as our unknown class because they are rare and difficult to find, and creating automated AI methods to search for these objects in new observations is crucial for inferring cosmological parameters.

We use the same labels for our source domain GZ2 SDSS dataset. For each class, we use galaxy IDs to find these objects in the GZ2 SDSS data and extract a similar number of examples, as was done in the DECaLS data. Finally, in both domains all classes have between 1--2.6 thousand images, except a much smaller cigar-shaped smooth class, with 334 images, which we purposefully do not augment, to test the performance of the model in the presence of a class imbalance. Finally, the source and target datasets each contain ${\approx}20,000$ images (GZ2 has $21784$, and DECaLS $19840$).

Both the SDSS and DECaLS data include three filter images ($i$,~$r$,~$g$). The SDSS DR7 data has a pixel scale of $0.396''$, while the DECaLS DR7 data has a scale of $0.262''$. Furthermore, the SDSS data includes galaxies with Petrosian half-light magnitude in the $r$-band $m_r<17.0$ (after the galactic extinction correction was applied). Deeper DECaLS images (with the lowest $r$-band magnitude of $m_r=23.6$, versus $m_r=22.2$ from SDSS) reveal spiral arms, weak bars, and tidal features not previously visible in SDSS imaging. The GZ3 DECaLS dataset was derived from SDSS DR8 imaging~\citep{AA2011}, so it only includes galaxies that are within both the DECaLS and SDSS DR8 footprint and have a slightly fainter magnitude limit of $m_r<17.77$. 

\subsubsection{Galaxy Zoo: SDSS Wide and Deep Field\\}
\label{sec:GZ2_SDSS}

Additionally, we use \textit{DeepAstroUDA} on a problem that includes different types of observational data from the same astronomical survey. The source domain includes wide-field SDSS data from GZ2, and the target domain comes from the deeper Stripe 82 region of SDSS, also available in GZ2~\citep{LS2008,LS2010,WL2013}.
Stripe 82 is a multiply-imaged section along the celestial equator in the Southern Galactic Cap, which results in better visibility of fainter objects. 

\setcounter{footnote}{0}
For the source domain data, we use the same 9-class wide-field dataset described in Section~\ref{sec:GZ2_SDSS_DECaLS}.
For our target data from Stripe 82, we use images from the co-added depth (set 2)~\footnote{Stripe 82 data is available at:\url{https://data.galaxyzoo.org}}.
These images are made from combining between 47--55 individual exposures, resulting in better detection of fainter features and improved seeing, compared to the source domain data.
This deeper dataset includes objects with magnitude limit $m_r<17.77$.
The images also include a modest color desaturation to deemphasize background noise in the co-added data.
In order to make both datasets as unique as possible, we allow a maximum of $5\%$ of extracted object overlap between the two datasets (objects found in a region covered by both wide and deep field observations).
We create the target domain dataset by extracting ${\approx}2,000$ images for each of the known $9$ classes and the unknown lens class, balancing the classes in the target dataset with a final size of ${\approx}20,000$ images.
These images did not pass any curation, so unlike the source domain, they also include examples that are harder to distinguish between classes.
We chose to do this to observe how \textit{DeepAstroUDA} performs in more realistic scenarios, where new data did not yet pass any human-level filtering.
The fact that this dataset includes harder to distinguish examples will also likely lead to more confusion in the crowd-sourced true labels, with possibly more incorrect true labels.

\subsection{Data Pre-Processing}
\label{sec:preprocess}

For all of our observational data, we use SExtractor~\citep{BA1996} to determine the center and radius of objects in the downloaded images, and then crop images to $256\times256$ pixels, using the extracted object properties to ensure no pertinent parts of the image have been inappropriately cut off. 

All datasets (simulated and observational) are normalized to have pixel values in $\left[0,1\right]$. Finally, the datasets are divided into training, validation, and test sets in proportions $60\%:20\%:20\%$.

\section{Results}
\label{sec:results}

In this section, we present the results of the \textit{DeepAstroUDA} model on three different types of DA problems that include observational data: 1) DA between different data releases of the same survey (LSST); 2) DA between two different astronomical surveys (SDSS and DECaLS); and 3) DA between two observing fields of different depths within the same survey (wide and deep fields of SDSS). In all three examples, we focus not only on the proper domain alignment and overlap of known classes, but also on adding an unknown class to all target domain datasets. This makes all of our examples harder open DA problems and also allows us to test anomaly detection (by discovering and clustering the unknown class samples). We show that our model is good at both aligning the known classes and discovering and clustering the unknown anomaly class, in all types of observational cross-dataset problems.

\subsection{Domain Adaptation Within a Survey: Different Data Releases of LSST}
\label{sec:within_survey_LSST_Y1Y10}

We first apply the \textit{DeepAstroUDA} model to the simplest example: two data releases from the same astronomical survey. In this case, the difference between the domains arises due to the different noise levels in the data (with subsequent data releases having lower noise levels due to the longer observation time). For this test, we used simulated mock LSST data (one and ten years of observations), described in detail in Section~\ref{sec:data_simulated}. The dataset includes three galaxy morphology classes: spiral, elliptical, and merging galaxies. As was shown already in~\cite{CK2021}, without DA, a model trained on one of these datasets cannot be used on the other one. In that paper, the model was trained on the low-noise Y10 data and applied to the noisier Y1 data. 

\begin{figure}
\begin{centering}
	\includegraphics[width=\linewidth]{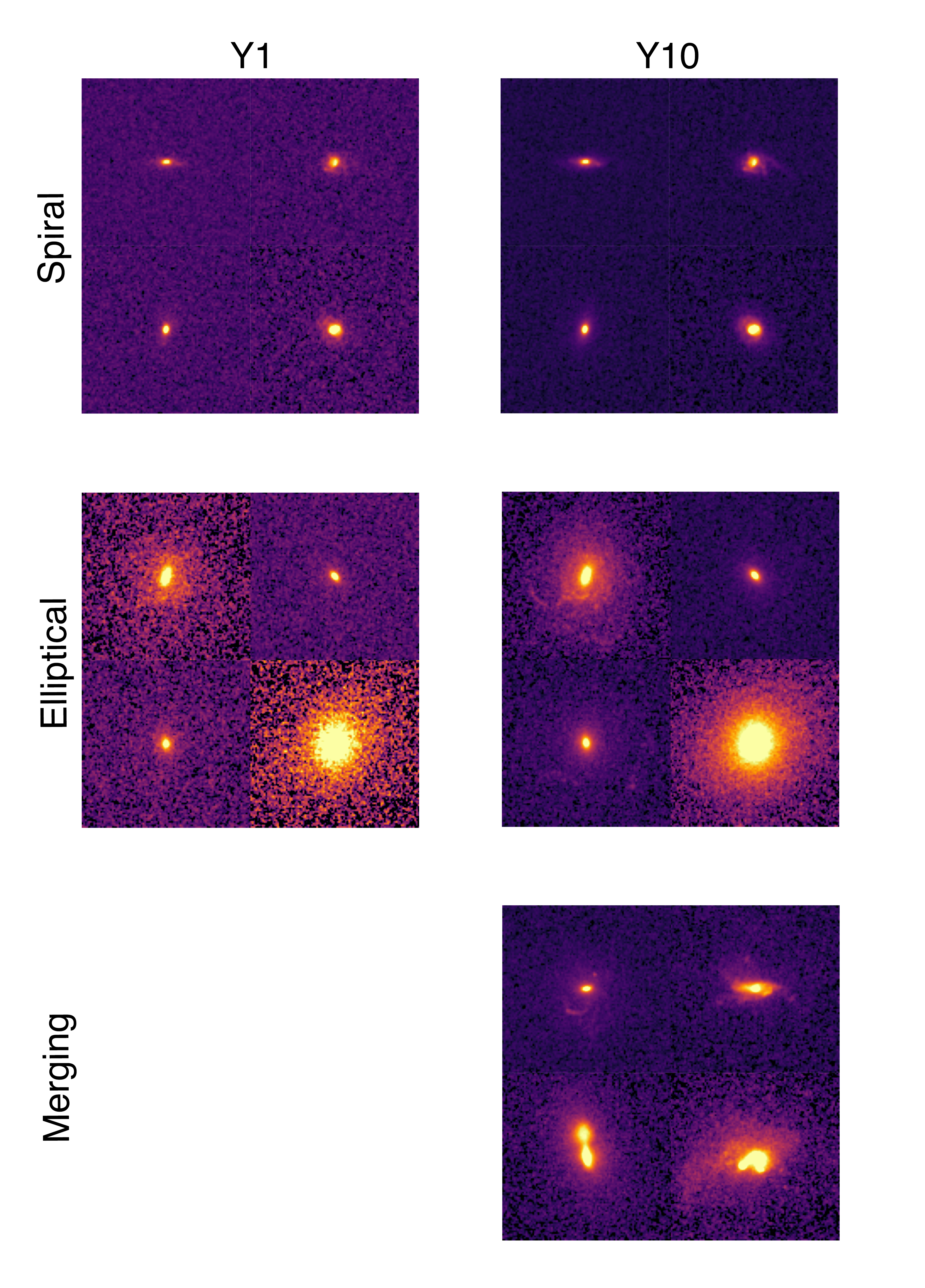}
    \caption{Example images from the source domain high-noise Y1 (left) and target domain low-noise Y10 (right) test set. The source domain contains two classes: spiral and elliptical galaxies, while the target domain also contain an unknown merging galaxies class.}
    \label{fig:example_3-class}
    \end{centering}
\end{figure}

In contrast to~\cite{CK2021}, here we focus mainly on a more realistic scenario, in which the model is trained on the first noisier data release (Y1), with the intention of later being applied to all other lower-noise data releases (for example Y10). In Figure~\ref{fig:example_3-class}, we show randomly selected example images from the source Y1 and target Y10 test data sets, denoted as High$\rightarrow$Low training. We additionally train another model on the reverse case (Low$\rightarrow$High), where the Y10 data is the source domain, as in~\cite{CK2021}. This facilitates a closer comparison to that paper. In this work, we consider a harder open DA problem in which spiral and elliptical galaxies are known classes present in both domains and merging galaxies are an additional unknown class present only in the target domain (while in~\cite{CK2021}, all three classes are present in both domains). In order to show how much the inclusion of DA helps with the performance in the target domain, we will compare the results of two models: 1) regular training on the source domain without any DA, and 2) training on the source domain with the inclusion of DA for the unlabeled target domain data. 

In Table~\ref{tab:lsst_results}, we show common performance metrics on the source and target test sets (accuracy, precision, recall, and F1 score, averaged over all classes present in each of the domains), with regular training, i.e., training with just ${\cal L}_{CE}$ on the source domain (top row), or training with DA (bottom row). All models in this paper were retrained $5$ times, with different random seeds used to initialize the weights, and we report the values of all performance metrics averaged across these runs. 

We can see that in both cases training without DA makes the model not work at all on the target domain data. Using DA, on the other hand, increases the accuracy in the target domain up to $74\%$. Adding DA leads to a substantial increase in the mean accuracy even in the source domain (by $9-12\%$), which ultimately reaches similarly high levels as the target domain accuracy (up to $76\%$). Furthermore, even though we are working on a harder problem, where mergers are an unknown target domain class, employing \textit{DeepAstroUDA} leads to $8\%$ better target domain accuracies compared to the results in~\cite{CK2021}. In addition, our methods produce more balanced performance across the source and target domains, considering accuracy, precision, and recall. By allowing the model to use more robust features, DA acts as a regularizer, making overfitting harder. Finally, by including active tuning of the DA loss parameters, the model can reach higher accuracies in fewer epochs (this example was impacted the most, with training stopping $15$ epoch earlier in the Low$\rightarrow$High case).

\begin{table}
   \centering
   \noindent\begin{minipage}[b]{0.99\columnwidth}
   \centering
    \caption{
    Mean performance metrics for \textit{ResNet50} when trained with regular training (top row) and with DA (bottom row). The top table shows results when low-noise data is used as the source domain and high-noise data as the target domain, while the bottom table shows results for the reversed case. The inclusion of DA increases the accuracy and other metrics for both source and target data, in both experiments.
    }
  \label{tab:lsst_results}
  \centering
  \begin{tabular}{|l || l |c c|}
\multicolumn{2}{c}{} &  \multicolumn{2}{c}{$\mathrm{Low}\rightarrow\mathrm{High}$}\\
 \hline Training      &   Metric   &  Source  & Target  \\\hline \hline
\multirow{4}{*}{Reg.}               &  Accuracy     &   $0.64 \pm 0.22$       &  $0.36 \pm 0.11$    \\ 
                                                &  Precision    &   $0.59 \pm 0.34$       &  $0.35 \pm 0.31$   \\
                                                &  Recall       &   $0.65 \pm 0.31$       &  $0.41 \pm 0.29$   \\
                                                &  F1 Score     &    $0.62 \pm 0.23$      &  $0.38 \pm 0.22$   \\\hline
\multirow{4}{*}{DA}         &  Accuracy     &   $0.76 \pm 0.12$       &  $0.74 \pm 0.09$   \\ 
                                                &  Precision    &   $0.74 \pm 0.13$       &   $0.71 \pm 0.19$   \\
                                                &  Recall       &   $0.75 \pm 0.08$       &   $0.74 \pm 0.21$     \\
                                                &  F1 Score     &   $0.74 \pm 0.08$       &   $0.72 \pm 0.32$  \\\hline
\multicolumn{4}{c}{}\\
\multicolumn{2}{c}{} & \multicolumn{2}{c}{$\mathrm{High}\rightarrow\mathrm{Low}$}\\
\hline Training      &   Metric   &  Source  & Target  \\\hline \hline
\multirow{4}{*}{Reg.}               &  Accuracy     &   $0.61 \pm 0.32$       &  $0.33 \pm 0.48$    \\ 
                                                &  Precision    &   $0.65 \pm 0.34$       &  $0.31 \pm 0.46$   \\
                                                &  Recall       &   $0.54 \pm 0.28$    &  $0.33 \pm 0.39$   \\
                                                &  F1 Score     &    $0.59 \pm 0.22$      &  $0.32 \pm 0.31$   \\\hline
\multirow{4}{*}{DA}         &  Accuracy     &   $0.70 \pm 0.20$         &  $0.73 \pm 0.18$   \\ 
                                                &  Precision    &   $0.64 \pm 0.30$       &   $0.72 \pm 0.17$   \\
                                                &  Recall       &   $0.66 \pm 0.24$       &   $0.74 \pm 0.20$     \\
                                                &  F1 Score     &   $0.65 \pm 0.19$       &   $0.73 \pm 0.13$  \\\hline
\end{tabular}
\end{minipage}
\end{table}

\begin{figure}
	\includegraphics[width=\linewidth]{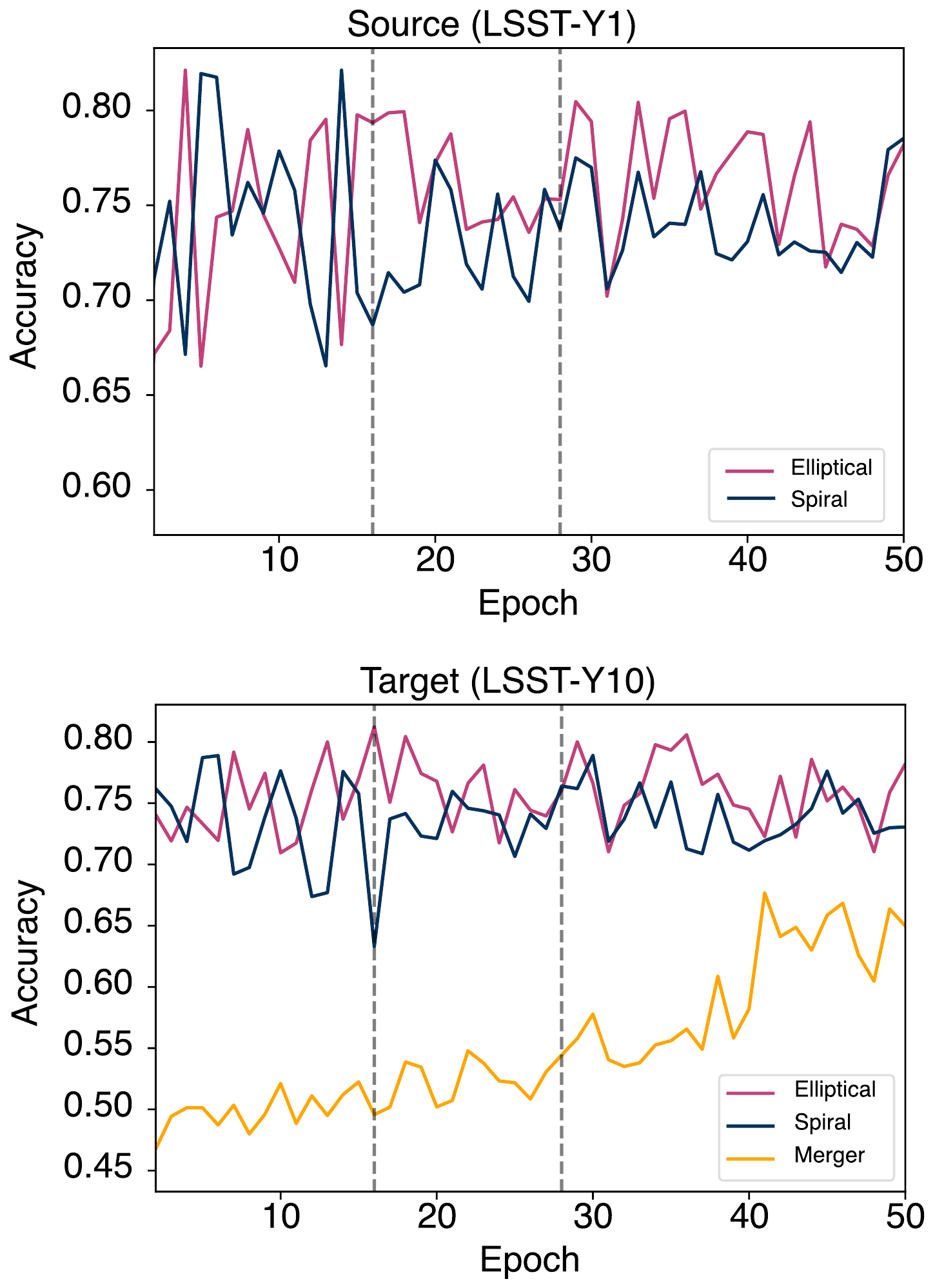}\\
    \caption{The source domain Y1 (top) and target domain Y10 (bottom) accuracies during model training. Elliptical (violet) and spiral galaxies (navy) are present in both domains, while merging galaxies (yellow) represent the unknown anomaly class, present only in the target domain. Vertical gray dashed lines show epochs in which the tuner changes the ES loss hyperparameters, which results in improved clustering and better performance on the anomaly class.}
    \label{fig:accuracies}
\end{figure}

\begin{figure}
	\includegraphics[width=\linewidth]{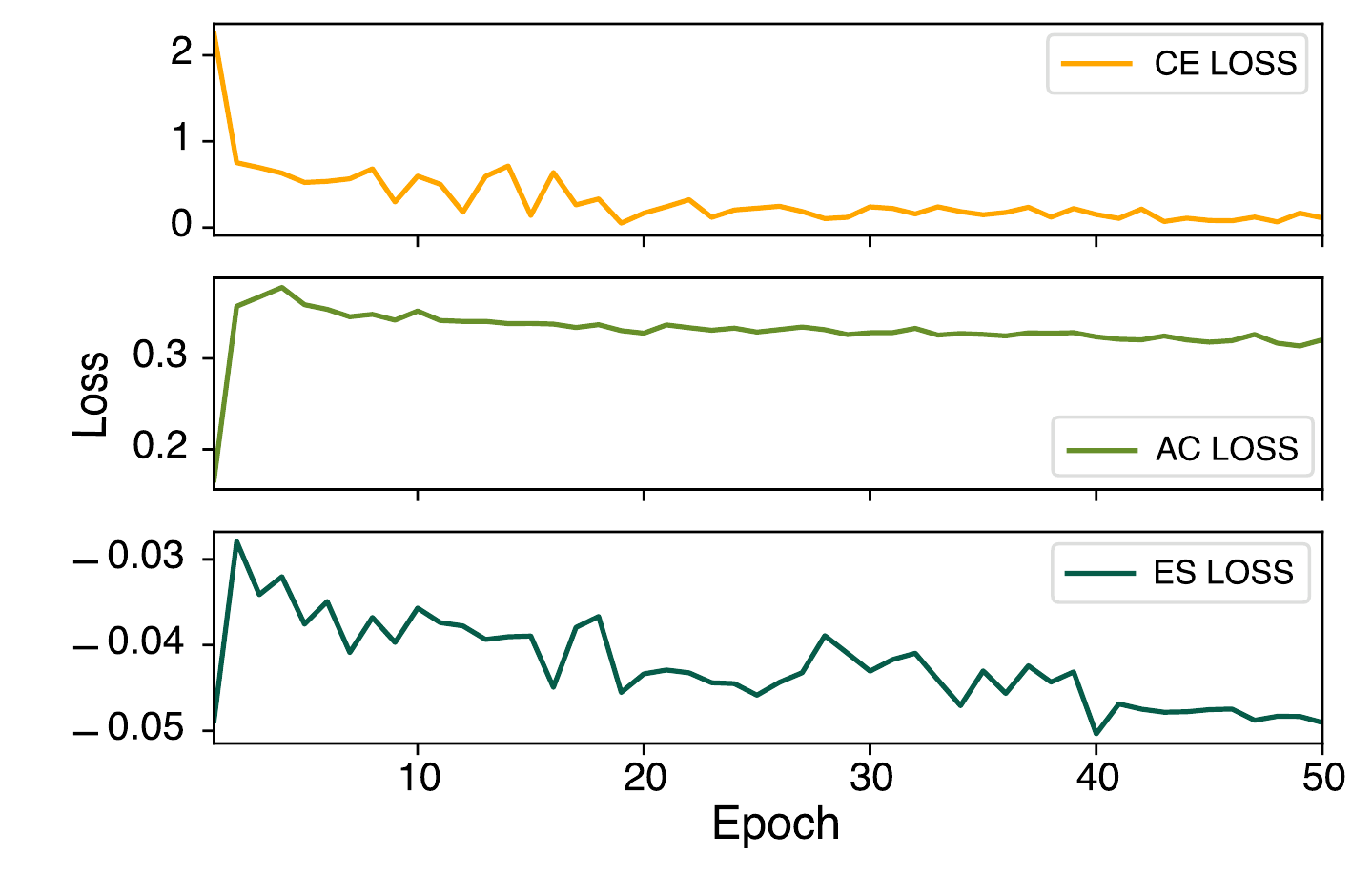}\\
    \caption{The loss function values during the $\mathrm{High}\rightarrow\mathrm{Low}$ training: CE loss (top, yellow), AC loss (middle, green), and ES loss (bottom, dark green).}
    \label{fig:losses}
\end{figure}

\begin{figure}
\centering
	\includegraphics[width=0.78\linewidth]{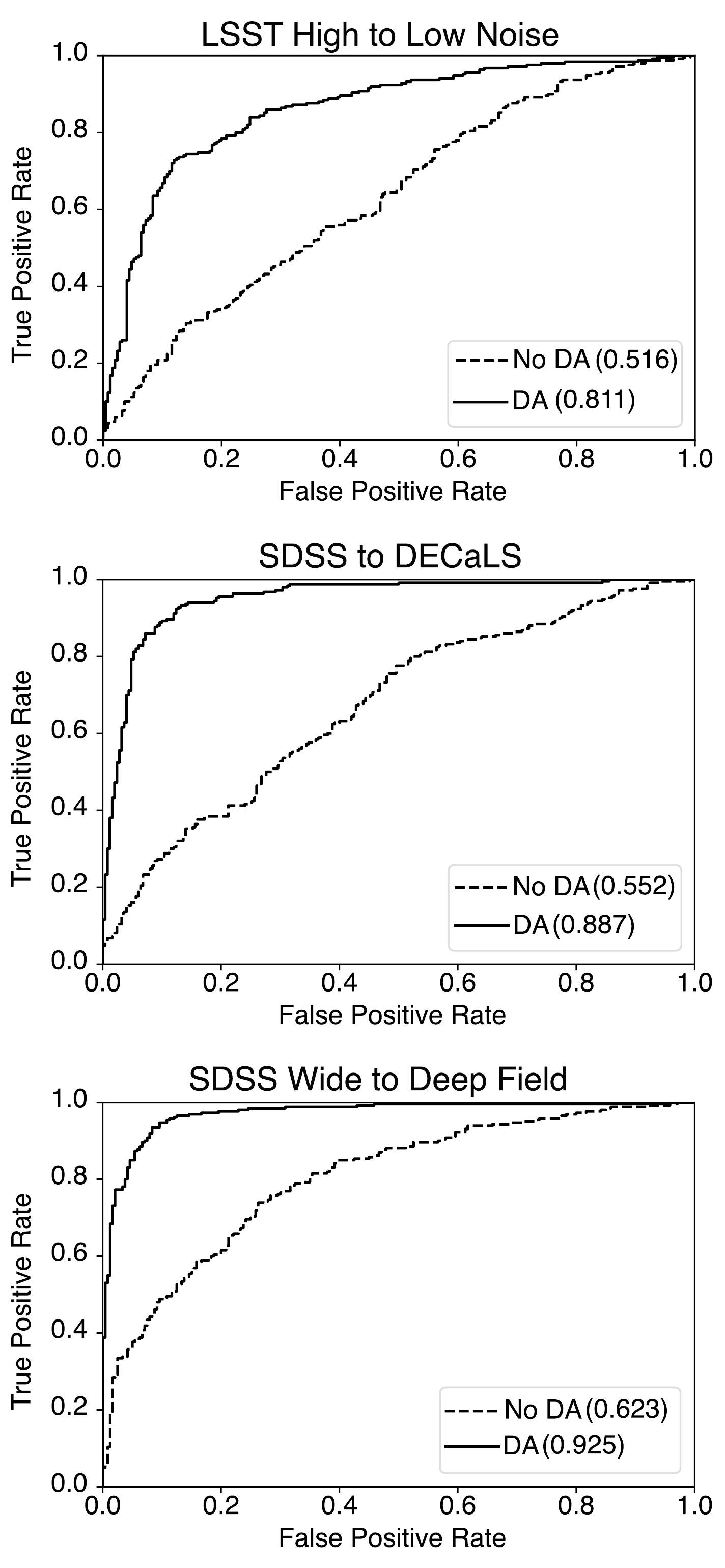}\\
    \caption{ROC curves and AUC scores (given in brackets) for models trained with regular training (dashed black lines) and models which include DA (solid black lines). We present results for target test set for more realistic $\mathrm{High}\rightarrow\mathrm{Low}$ noise experiment within one survey (top), cross-survey experiment (middle), and wide and deep wield within one survey experiment (bottom). DA substantially improves performance in all three experiments.} 
    \label{fig:ROC}
\end{figure}

To better illustrate the effects of DA training, in Figure~\ref{fig:accuracies} we show how the accuracies for all three classes change during the more realistic High$\rightarrow$ Low noise training case. Vertical dashed lines mark epochs where our hyperparameter tuner changes the ES loss parameters to try to improve clustering. We also show how all three loss functions change during training in Figure~\ref{fig:losses}. Furthermore, for the more realistic $\mathrm{High}\rightarrow\mathrm{Low}$ experiment, we  also present the Receiver Operating Characteristic (ROC) curves (which shows the trade-off between the true positive rate and the false-positive rate of a classifier) on the target test set, for models trained with regular and DA training, and report their Area Under the Curve (AUC) scores (top plot of Figure~\ref{fig:ROC}).

\begin{figure}\begin{centering}
	\includegraphics[width=0.9\linewidth]{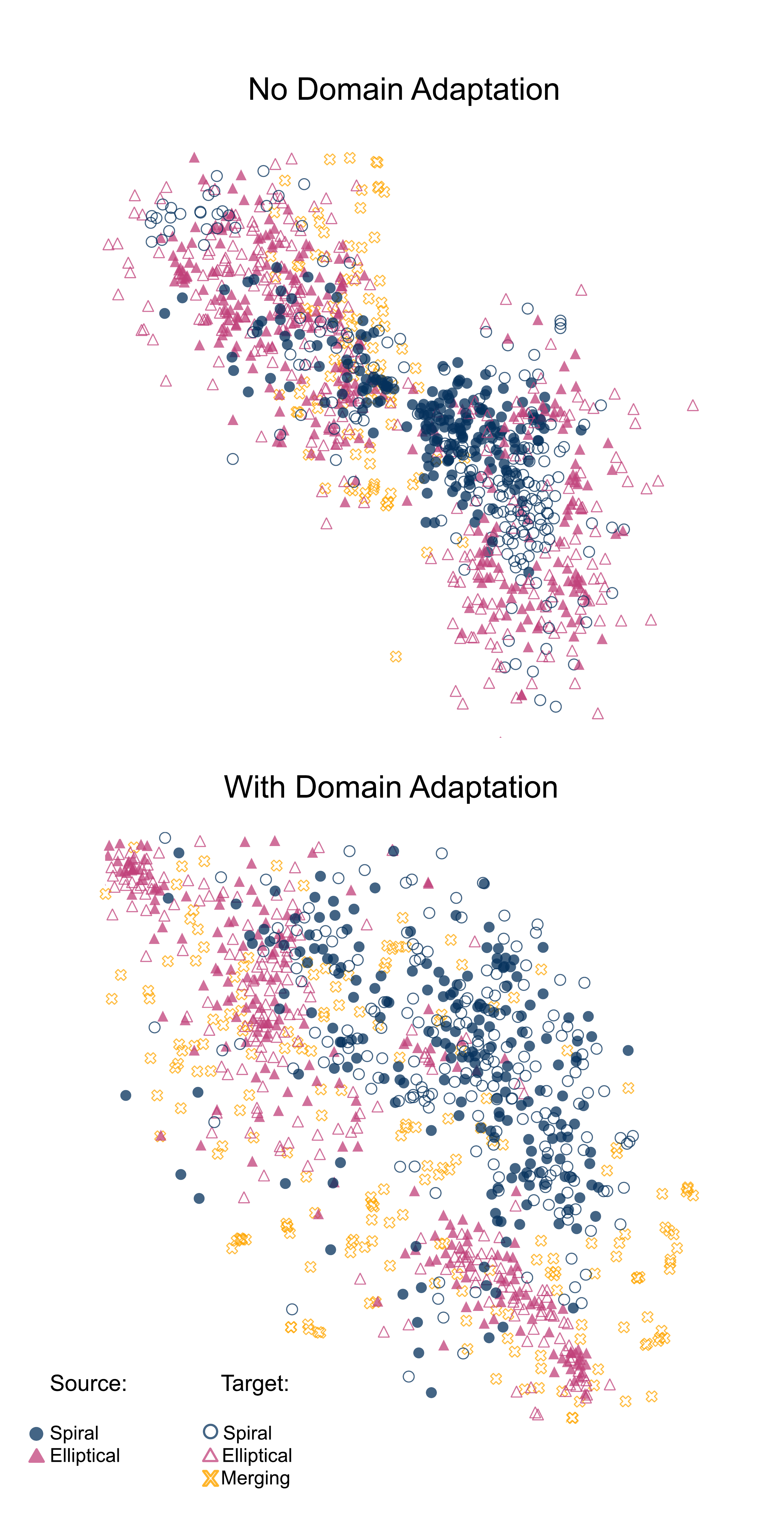}\\
    \caption{The t-SNE plots when training with Y1 data as the source domain and Y10 data as the target domain ($\mathrm{High}\rightarrow\mathrm{Low}$). The top plot shows the latent space of the model trained without DA, while the bottom plot shows the model trained with DA. Using DA aligns known classes much better, but also allows the unknown class to be pushed to the outskirts.}
    \label{fig:tsne_lsst}
    \end{centering}
\end{figure}

Finally, to better understand how the inclusion of DA influences the way the model represents the data in its latent space, we show the t-SNE plot (of the isomap of the latent space representation of the data) in Figure~\ref{fig:tsne_lsst}. We can see that without DA (top), the domains do not overlap very well. In this example, the domains are quite similar (both datasets include the same galaxy images, and the only difference is the noise levels), so the model trained without DA is able to place some target-domain images correctly, but the unknown merger class is completely overlapping with the known classes. The inclusion of DA (bottom) makes the known classes overlap much better, while the unknown target class is pushed away from the known data.

\subsection{Domain adaptation across surveys: SDSS to DECaLS}
\label{sec:across_surveys}

As mentioned above, DA in astronomy has only been tried between simulated mock data mimicking a particular telescope and real observations from the same telescope~\cite{CK2021MNRAS}. The success of the methods used so far strongly depended on the minimization of the dataset shift, by making simulated mock data as similar to real data as possible, before any model training was performed. The difference between data from two different astronomical surveys can, on the other hand, be much larger: from different noise levels, PSF blurring, pixel scale, survey depth, etc.
Here, we present the first successful cross-survey DA result in astronomy.  We will focus on a more complicated 10-class galaxy morphology problem that includes data from two different astronomical surveys: SDSS and DECaLS. These more complex dataset shifts require more flexible DA methods. If two data distributions look very different, or do not share all of the same classes, simpler methods that explicitly minimize a distance metric between the latent data distributions and are not class-aware might not be powerful enough to properly align the two data distributions. For that reason, in this work we develop a much more flexible semi-supervised method that is able to align similar samples for any kind of data distribution shapes and possible overlaps.

In Table~\ref{tab:10_results}, we again report the mean accuracy, precision, recall, and F1 score for the source and target test sets: regular training (top row), and training with DA (bottom row). Again, the use of DA improves performance in both data domains, from a $5\%$ increase in accuracy in the source domain to a $36\%$ increase in accuracy in the target domain. Furthermore, the performance and behavior of the model (reflected in all performance metrics) is similar across the domains after the inclusion of DA. Also, the model performance reaches levels similar to our first, much simpler example in Section~\ref{sec:within_survey_LSST_Y1Y10}, even though this problem has a much larger distribution shift and more classes. In Figure~\ref{fig:ROC} (middle panel) we show the ROC curves and AUC scores on the target test set for models using regular or DA training.

\begin{table}
   \centering
   \noindent\begin{minipage}{\linewidth}
   \centering
    \caption{Mean performance metrics for \textit{ResNet50} on SDSS (source) and DECaLS (target) test data for regular training without DA (top row) and training with DA (bottom row). The inclusion of DA increases accuracy and other metrics for both the source and target data.
    }
  \label{tab:10_results}
  \centering
  \begin{tabular}{|l || l |c c|}
\multicolumn{2}{c}{} &  \multicolumn{2}{c}{$\mathrm{SDSS}\rightarrow\mathrm{DECaLS}$}\\
 \hline Training      &   Metric   &  Source  & Target  \\\hline \hline
\multirow{4}{*}{Reg.}               &  Accuracy     &   $0.77 \pm 0.20$       &  $0.43 \pm 0.32$    \\ 
                                                &  Precision    &   $0.52 \pm 0.23$       &  $0.42 \pm 0.43$   \\
                                                &  Recall       &   $0.70 \pm 0.25$       &  $0.33 \pm 0.30$   \\
                                                &  F1 Score     &    $0.60 \pm 0.18$      &  $0.37 \pm 0.25$   \\\hline
\multirow{4}{*}{DA}         &  Accuracy     &   $0.82 \pm 0.09$       &  $0.79 \pm 0.10$   \\ 
                                                &  Precision    &   $0.79 \pm 0.06$       &   $0.73 \pm 0.21$   \\
                                                &  Recall       &   $0.84 \pm 0.13$       &   $0.81 \pm 0.18$     \\
                                                &  F1 Score     &   $0.81 \pm 0.07$       &   $0.77 \pm 0.14$  \\\hline
\end{tabular}
\end{minipage}
\end{table}

\begin{figure}
    \centering	
    \includegraphics[width=\linewidth]{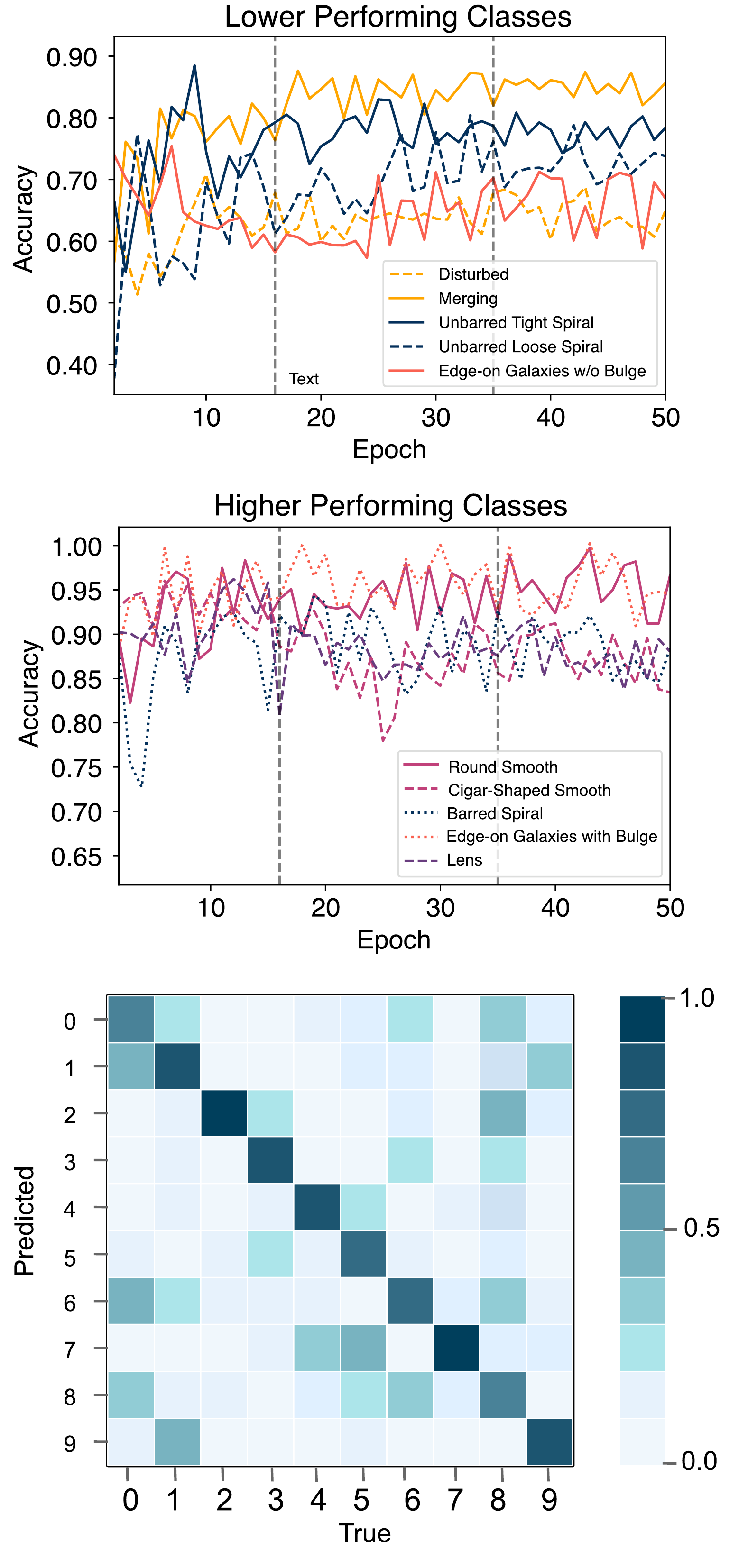}\\
    \caption{Individual class accuracies in the target domain during training. The top figure shows lower performing classes, while the middle figure shows higher performing classes. Finally, the bottom figure shows the confusion matrix of the final trained model, when applied to the target domain test set (disturbed (0), merging (1), round smooth (2), cigar-shaped smooth (3), barred spiral (4), unbarred tight spiral (5), unbarred loose spiral (6), edge-on without bulge (7), edge-on with bulge (8), lenses (9)). The performance on all classes is very good, with slight confusion between several classes that indeed look visually similar, for example disturbed and merging galaxies.}
    \label{fig:per_class}
\end{figure}

To further illustrate the performance of the model on each of the classes, in Figure~\ref{fig:per_class} we show the individual target domain class accuracies and how they change during training. For better readability, we separate classes into lower performing (top figure) and higher performing (middle figure) classes. Additionally, we present the confusion matrix (bottom figure), to further illustrate the confusion between similar classes when the final trained model is applied to our target data test set of images. We can clearly see that the model confuses merging and disturbed galaxies, which can both exhibit asymmetric disturbed morphology and are both associated with galaxy interaction processes. Also, barred spiral and unbarred tight spiral are often incorrectly classified as the class edge-on without bulge, which also makes sense, given that all three classes actually include spiral galaxies with different orientations. Similar conclusions can be made for unbarred tight spiral and unbarred loose spiral, which are often confused with the class edge-on with bulge. We have also noticed some occurrence of errors in which a spiral galaxy that is very small with hard-to-distinguish features gets classified as one of the smooth classes.

\begin{figure}
\centering
	\includegraphics[width=0.9\linewidth]{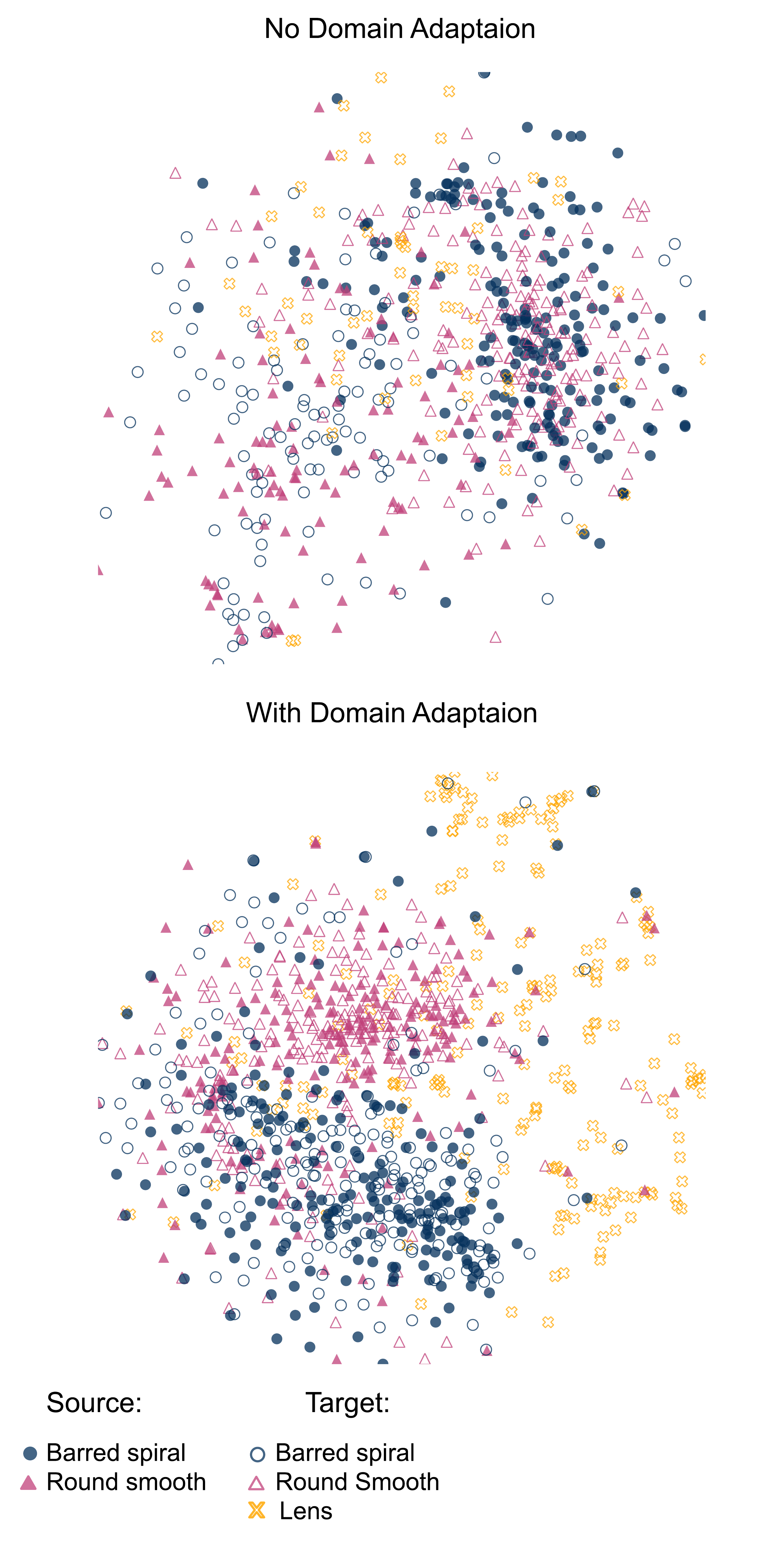}\\
    \caption{The t-SNE plots when training on 10-class SDSS data (source) and testing on DECaLS data (target). We choose to plot 3 classes for better visibility: round smooth (class 2), barred spiral (class 4), and lens (class 9) anomaly class in the target domain. Again, training with DA (bottom plot) helps correctly overlap the known classes and push away the anomaly class.}
    \label{fig:tsne_SDSS_DECaLS}
\end{figure}

Figure~\ref{fig:tsne_SDSS_DECaLS} shows t-SNE plots from the model with regular training and training with DA. For better visibility, we choose to plot only three out of ten classes: round smooth, barred spiral, and the unknown gravitational lens class. In this more complicated and larger domain shift problem, the model trained using regular training on the source domain data tends to flip classes from the two domain, overlapping the barred spiral class from the source domain with the round smooth class from the target and vice versa. The unknown lens class is also completely overlapping with known classes. The inclusion of DA during training corrects this behavior and correctly aligns the known classes, while pushing the unknown class to the outskirts.

\begin{figure*}
\begin{centering}
	\includegraphics[width=0.8\linewidth]{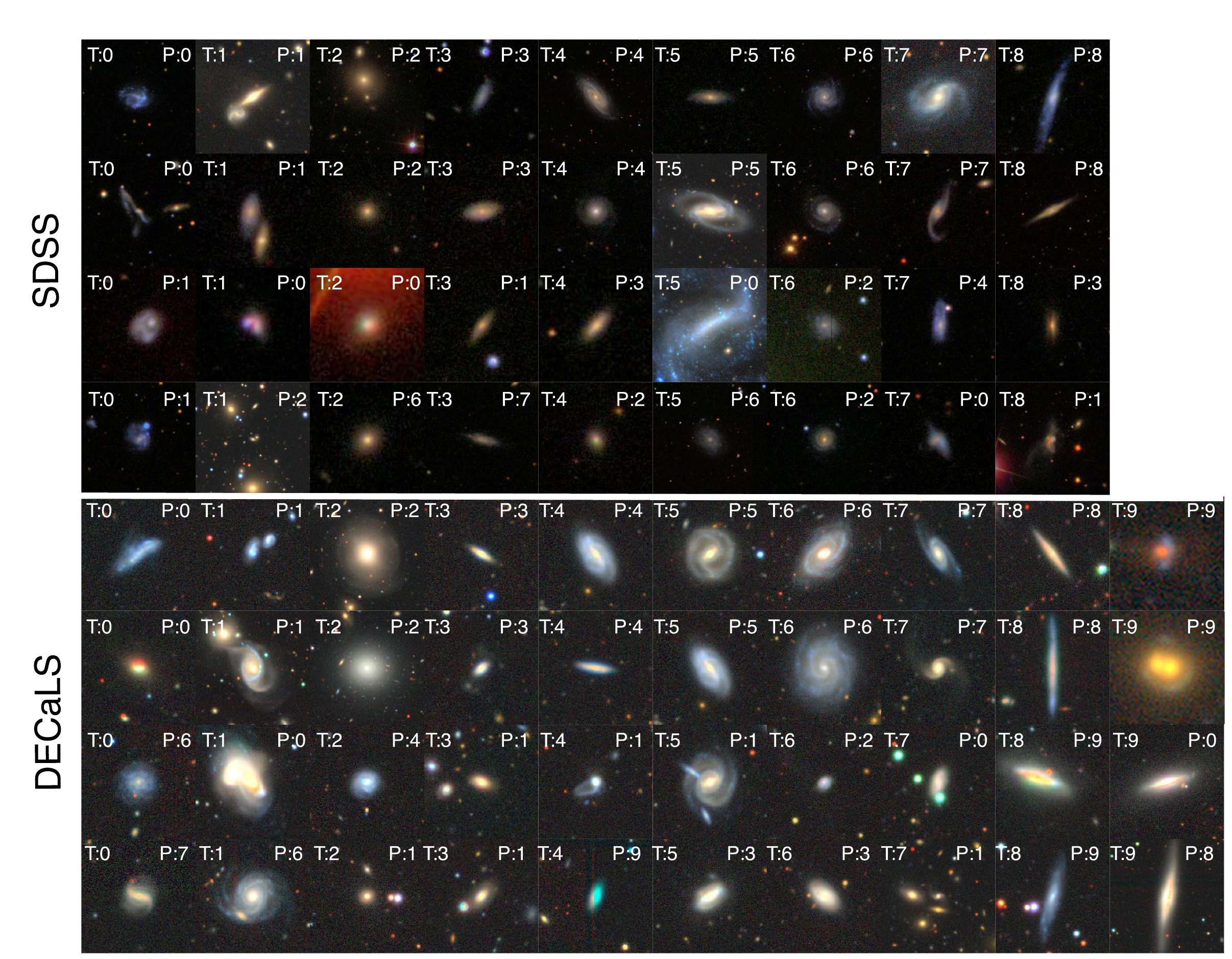}\\
    \caption{Example images from source SDSS data (top) and target DECaLS data (bottom). Each image contains the true label in the top left corner and the predicted label in the top right corner. For each domain, we give most often correctly classified examples in the top two rows and most often incorrectly classified examples in the bottom two rows. Classes are: disturbed (0), merging (1), round smooth (2), cigar-shaped smooth (3), barred spiral (4), unbarred tight spiral (5), unbarred loose spiral (6), edge-on without bulge (7), edge-on with bulge (8), lenses (9).}
    \label{fig:example-10-class}
    \end{centering}
\end{figure*}

\begin{figure*}
\begin{centering}
	\includegraphics[width=0.8\linewidth]{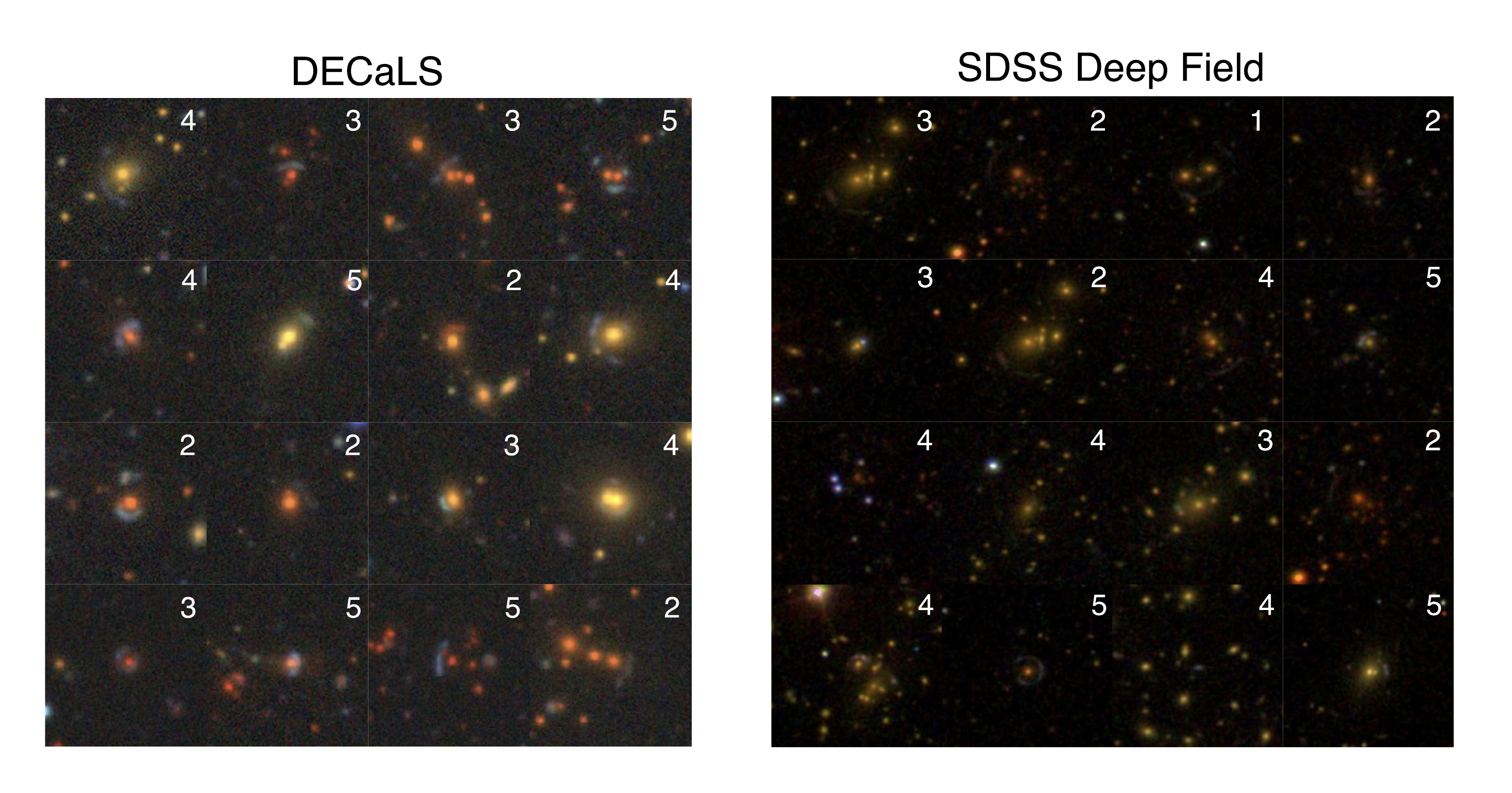}\\
    \caption{Randomly selected example images of correctly classified lenses from DECaLS data (left) and SDSS Stripe 82 deep field (right). Each image contains the number of models (out of $5$, trained with different random seed initializations) that correctly classified the example.}
    \label{fig:lens}
    \end{centering}
\end{figure*}

In Figure~\ref{fig:example-10-class}, we show example images from the source SDSS (top) and target DECaLS (bottom) test data sets. For each of the domains, the top two rows show examples of images most often correctly classified, while the bottom two rows show examples of most often incorrectly classified (considering the $5$ models trained with different random seed initializations). All images show the true class in the top left corner and predicted class in the top right corner, for one of our trained models. In Figure~\ref{fig:lens} (left), we also plot examples of correctly classified lenses in DECaLS data. Each image also includes the number of models (out of $5$) that were able to discover the lens. 

In Figure~\ref{fig:mean_acc} (left plot), we show how the accuracies of the target domain change during training. We present the mean accuracy of the known $9$ classes and the unknown class (gravitational lens) accuracy separately. We can see that the known classes reach quite high accuracy, around $80\%$, which is very good given the fact that this dataset includes several morphologically or visually similar classes (for example, disturbed and merging galaxies, or cigar-shaped smooth and edge-on without bulge). Furthermore, the model is even better at finding and classifying the unknown anomaly class, with the lens class reaching an accuracy of almost $90\%$, proving that this DA method can be used successfully not only to bridge the gap between different observational datasets, but also to search for unknown objects of interest, such as gravitational lenses.

When examining images from our datasets, we noticed examples in which true labels from crowd sourcing do not seem to be correct. In some of these cases, we find the predicted labels to be more appropriate, which makes us believe that the model had enough correctly labeled samples to properly understand the galaxy morphology and correctly classify even those examples that include questionable true labels. For example, in Figure~\ref{fig:example-10-class} the last image in the bottom row has a true lens label, while the network predicts class edge-on with bulge, which upon visual inspection seem like a more appropriate class. More evidence for this conclusion can be seen in the confusion matrix, where the biggest confusion between classes occurs for truly morphologically similar classes. 

\subsection{Domain adaptation within a survey: wide and deep fields of SDSS}
\label{sec:within_survey_SDSS_wide_deep}
Finally, DA can help bridge the gap between data from wide and deep fields within the same survey. Here, we focus on the wide field (source data) and Stripe 82 deep field (target data) of SDSS. We use the same 10-class problem as in the previous cross-survey example. As we describe in Section~\ref{sec:GZ2_SDSS}, we use the same source dataset as in the cross-survey example, and we make the target domain using randomly chosen examples from Stripe 82, which also contains crowd-sourced GZ2 labels. Compared to the previous cross-survey example, where we use curated datasets that contain cleaner examples, the target domain Stripe 82 data in this test is not curated and can include galaxies much harder to correctly classify into one of the ten classes. This makes classification using AI harder, but will also make human labeling more difficult, probably containing an even larger percentage of incorrect true labels.

In Table~\ref{tab:10_results_stripe}, we again present the performance metrics for the models trained with and without DA, when applied to the source and target test datasets. As before, training without DA makes the model useless on target data, whereas the inclusion of DA allows the model to perform well on both wide and deep fields of the same survey, with similar performance metrics across domains. The mean accuracy with DA reaches $81\%$ in the target domain and $84\%$ in the source domain, which is even slightly higher than in our previous cross-survey example. In Figure~\ref{fig:ROC} (bottom panel) we show ROC curves and AUC scores on the target test set for models trained with regular and with DA training.

\begin{figure*}[!htb]
\begin{centering}
	\includegraphics[width=0.86\linewidth]{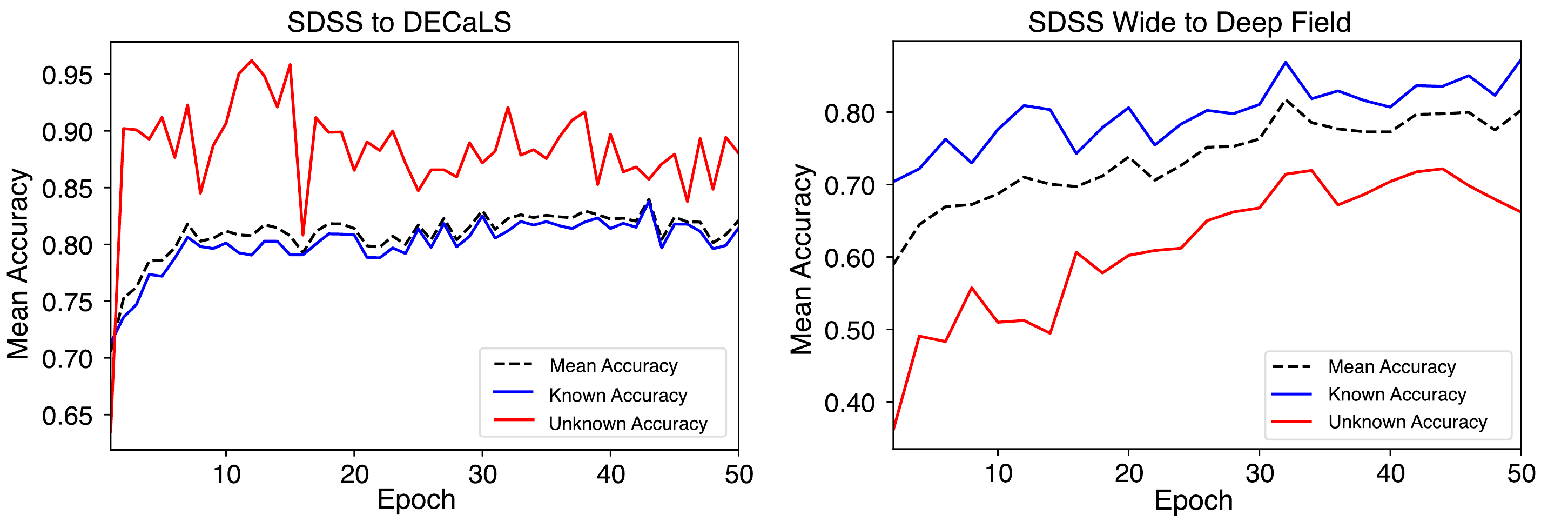}\\
    \caption{The target domain mean known (blue) and unknown class (red) accuracies during training. The mean accuracy for all classes is plotted with a dashed black line. The left plot shows results for the cross-survey problem (SDSS to DECaLS), where the target domain includes well curated examples, with clearer class distinctions. The right plot shows DA from wide to deep field of SDSS, where the target domain also contains harder examples. This leads to a lower increase in performance on the unknown lens class, which often gets confused with the edge-on with bulge class. Still, the increase in performance during model training is evident and the final performance on the unknown class is much better than for the model trained without any DA (since that model does not work at all on both known classes and the unknown class in the target domain).}
    \label{fig:mean_acc}
    \end{centering}
\end{figure*}

We omit plotting the confusion matrix for this example, since it looks very similar to the cross-survey example shown in Figure~\ref{fig:per_class}. Again, the biggest confusions occur between disturbed and merging galaxies (classes $0$ and $1$) and between different sub-classes of spiral galaxies (classes $4$, $5$ and $6$ incorrectly classified as $7$ or $8$). The most notable difference, in this example, is the presence of the somewhat unexpected confusion between the class edge-on with bulge (class $8$) and the unknown gravitational lens class (class $9$). This probably leads to lower performance on the unknown class, which we can see in Figure~\ref{fig:mean_acc} (right), showing the mean accuracy for all known classes and the accuracy of the unknown class in the target domain during model training.  As in the cross-survey example, the mean known class accuracy reaches levels around $80\%$, but the unknown class accuracy is in this case somewhat lower. After examining lens images from the SDSS deep field, we concluded that the lenses indeed look less prominent compared to the DECaLS data, which we believe has led to both more incorrect true labels from crowd-sourcing, and more confusion in the predictions by the \textit{DeepAstroUDA} model.

\begin{table}
   \centering
   \noindent\begin{minipage}{\linewidth}
   \centering
    \caption{Performance metrics for \textit{ResNet50} on SDSS wide field (source) and SDSS Stripe 82 deep field (target) test data for regular training without DA (top row) and training with DA (bottom row). The inclusion of DA increases the accuracy for both source and target data.
    }
  \label{tab:10_results_stripe}
  \centering
  \begin{tabular}{|l || l |c c|}
\multicolumn{2}{c}{} &  \multicolumn{2}{c}{$\mathrm{Wide}\rightarrow\mathrm{Deep}$}\\
 \hline Training      &   Metric   &  Source  & Target  \\\hline \hline
\multirow{4}{*}{Reg.}               &  Accuracy     &   $0.76 \pm 0.22$       &  $0.43 \pm 0.37$    \\ 
                                                &  Precision    &   $0.68 \pm 0.12$       &  $0.42 \pm 0.22$   \\
                                                &  Recall       &   $0.76 \pm 0.21$       &  $0.44 \pm 0.32$   \\
                                                &  F1 Score     &    $0.72 \pm 0.12$      &  $0.43 \pm 0.19$   \\\hline
\multirow{4}{*}{DA}         &  Accuracy     &   $0.84 \pm 0.08$       &  $0.81 \pm 0.12$   \\ 
                                                &  Precision    &   $0.83 \pm 0.15$       &   $0.80 \pm 0.07$   \\
                                                &  Recall       &   $0.79 \pm 0.12$       &   $0.81 \pm 0.07$     \\
                                                &  F1 Score     &   $0.81 \pm 0.10$       &   $0.80 \pm 0.05$  \\\hline
\end{tabular}
\end{minipage}
\end{table}

\begin{figure*}
\begin{centering}
	\includegraphics[width=0.85\linewidth]{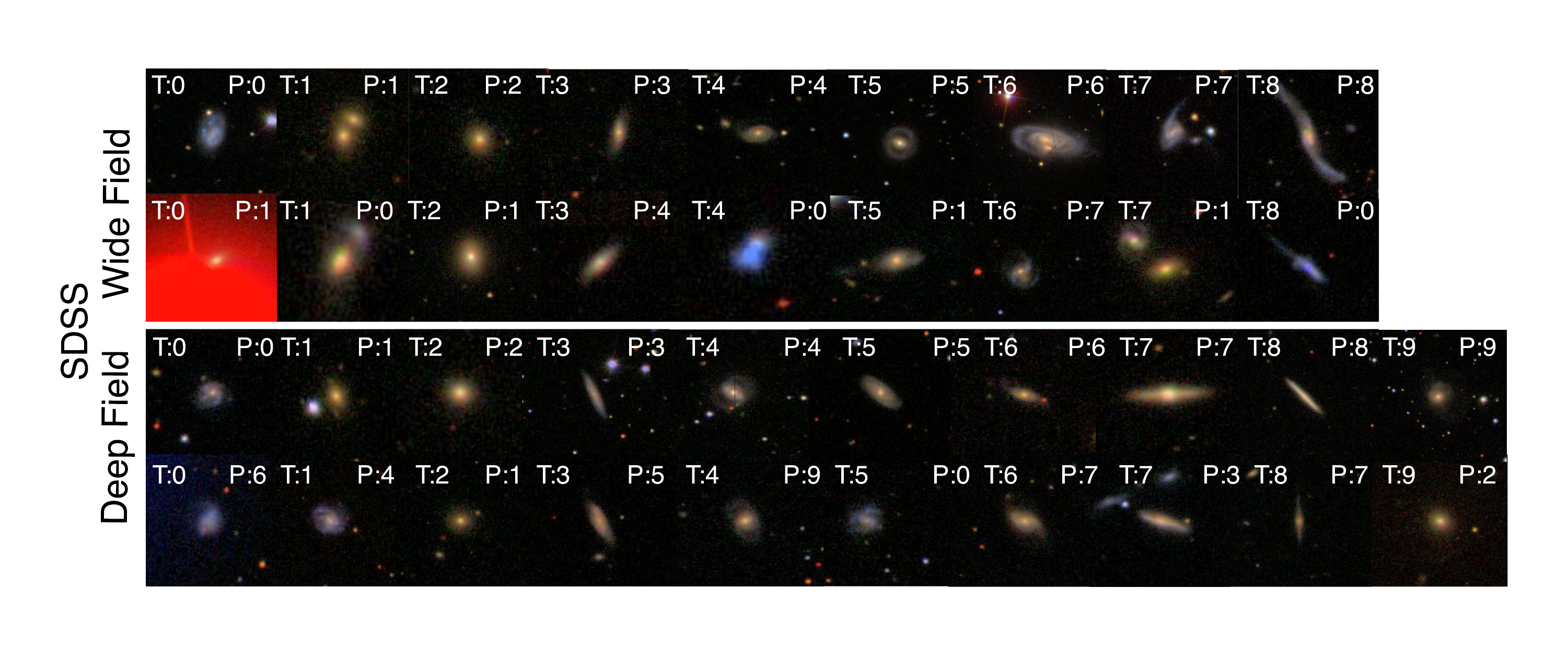}\\
    \caption{Example images from the source SDSS wide field (top) and the target SDSS Stripe 82 deep field (bottom). Each image contains the true label in the top left corner and the predicted label in the top right corner. For each domain, we give most often correctly classified examples in the top row and most often incorrectly classified examples in the bottom row. Classes are again: disturbed (0), merging (1), round smooth (2), cigar-shaped smooth (3), barred spiral (4), unbarred tight spiral (5), unbarred loose spiral (6), edge-on without bulge (7), edge-on with bulge (8), lenses (9).}
    \label{fig:examples_10-class_stripe}
    \end{centering}
\end{figure*}

In Figure~\ref{fig:examples_10-class_stripe}, we again show examples most often correctly classified and incorrectly classified in both domains, the wide field (top) and the target Stripe 82 deep field of SDSS data (bottom), as well as their true and predicted classes. In Figure~\ref{fig:lens} (right), we show examples of discovered lenses, and give the number of times the models correctly classified each of the lenses (out of $5$ models trained with different random seed initializations).

\subsection{Comparison to other methods and benchmarks}
\label{sec:benchmark}

To further illustrate the power of our method compared to similar UDA methods, capable of handling any type of dataset overlap, we compare the results of the \textit{DeepAstroUDA} and DANCE methods~\citep{SK2020} on both the Office dataset~\citep{SK2010}, which is a standard benchmarking dataset, and our LSST mock datasets described in Section~\ref{sec:data_simulated}. In Table~\ref{tab:comparison}, we show the target test set accuracies for \textit{DeepAstroUDA} and DANCE methods on Office and LSST data ($\mathrm{Low}\rightarrow\mathrm{High}$ and $\mathrm{High}\rightarrow\mathrm{Low}$ experiments).

\textit{DeepAstroUDA} outperforms DANCE on both Office and LSST mock data. Due to the active hyperparameter tuning, our model is much easier to train, because it converges on well performing hyperparameters and evolves them as the training progresses. On the other hand, the DANCE method is much harder to train since well performing parameters need to be found manually, and they are kept constant throughout the training. Finally, DANCE results on the Office dataset presented here are lower than the values reported in~\cite{SK2020}. We were not able to replicate their reported results using the hyperparameters authors list in their paper. After searching for better performing values we were able to improve DANCE model performance, but the highest accuracies that we report here are still lower than in the original paper.

\begin{table}
   \centering
   \noindent\begin{minipage}{\linewidth}
   \centering
    \caption{Performance comparison for \textit{ResNet50} (results from a single trained model only) with DA included in the training via \textit{DeepAstroUDA} and DANCE methods. We show the target test set accuracy for \textit{DeepAstroUDA} and DANCE methods on Office and LSST data ($\mathrm{Low}\rightarrow\mathrm{High}$ and $\mathrm{High}\rightarrow\mathrm{Low}$ experiments).
    }
  \label{tab:comparison}
  \centering
  \begin{tabular}{|l || l |c|}
 \hline Training      &   Dataset     & Target  \\\hline \hline
\multirow{4}{*}{DANCE}               &  Office           &  $0.82$    \\ 
                                                & $\mathrm{Low}\rightarrow\mathrm{High}$             &  $0.37$   \\
                                                &  $\mathrm{High}\rightarrow\mathrm{Low}$              &  $0.46$   \\\hline
\multirow{4}{*}{\textit{DeepAstroUDA}}         &  Office     &     $\bm{0.91}$   \\ 
                                                &  $\mathrm{Low}\rightarrow\mathrm{High}$             &   $\bm{0.74}$   \\
                                                & $\mathrm{High}\rightarrow\mathrm{Low}$              &   $\bm{0.73}$  \\\hline
\end{tabular}
\end{minipage}
\end{table}

\section{Discussion and Conclusion}
\label{sec:conclusion}

\textit{DeepAstroUDA} is a flexible semi-supervised DA algorithm that can handle any kind of cross-dataset problem. It can be used for classification, regression, and anomaly detection, and it works well even in the presence of non-overlapping (or unknown) classes in any of the two data domains. This means that it can be used for open, partial, and mixes of open and partial DA problems. This is a crucial feature for successful implementation of DA algorithms in real scientific applications, where we often will not have well curated datasets. 

The inclusion of DA in astronomy and cosmology is a necessity, as most types of studies include the use of multiple simulated and observational datasets. In this work, we address open DA problems related to galaxy morphology classification. We focus on observational data and demonstrate the use of the method in three different scenarios: 1) data collected after different numbers of observing years, i.e., different data releases of the same survey; 2) data from two different surveys; and 3) data from wide and deep fields of the same survey. These three scenarios will have different combinations of factors that make the datasets different: pixel scale, noise levels, PSF blurring, the depth of the survey, and the magnitude limit of the objects that can be observed. All the results presented in this paper show that \textit{DeepAstroUDA}: 

\begin{enumerate}
    \item can successfully be used on difficult domain shift problems that include multiple observational survey datasets;
    \item allows the trained model to perform well even on the unlabeled target domain data, with an observed increase in accuracy up to $40\%$;
    \item handles any type of domain overlap and performs even in the presence of unknown classes, which can be used for anomaly detection tasks, like searching for merging galaxies or gravitational lenses;
    \item increases accuracy on both source and target data, and makes performance of the model consistent across both domains;
    \item is easy to train because it includes active hyperparameter tuning, which allows the method to converge to optimal loss parameter values (even if the initial guess is not good) and continuously evolve them as the training progresses and the latent data distributions evolve.
\end{enumerate}

In this work, we use Galaxy Zoo labels, made by thousands of volunteers. This process naturally makes true labels prone to errors since it is often very hard to visually distinguish between different morphological classes. Still, most true labels should be correct, since all of the models we train seemed to be able to correctly learn the distinction between galaxy classes. For example, upon closer visual inspection of examples from our test set, we noticed images for which we actually agree more with the (incorrect) predicted class given by the model than with the true crowd-sourced label. Furthermore, the biggest confusion between classes occurs for truly hard to distinguish objects, like disturbed and merging galaxies, where both classes contain very asymmetric objects, with tidal tails and other interesting features, which are a product of galaxy interactions.

An important feature of \textit{DeepAstroUDA} is its ability to be used as an anomaly detection algorithm in the unlabeled target domain. We focus here on merging and gravitationally lensed galaxies. Merging galaxies are crucial for understanding the process of galaxy evolution, formation of different galaxy morphologies, and star formation. Unfortunately, they are very difficult to find in real photometric observations. Without spectroscopic observations, we often cannot be sure if a pair of galaxies is truly merging or if it is just visually overlapping. With future surveys like LSST, where we expect that over $60\%$ of objects will be overlapping, discovering true mergers will be even more difficult. Methods such as \textit{DeepAstroUDA} will be a necessity to allow us to use the information available in other simulated and real datasets to search for merging galaxies in new observations.

Strong gravitational lensing of galaxies (as well as supernovae and quasars) is an important cosmological probe. Future surveys like LSST will discover unprecedented numbers of lensed objects, and efficient discovery and creation of complete catalogs will allow us to put tighter constraints on cosmological parameters and better understand both dark energy and dark matter. Regular AI models trained on simulated samples of strong gravitation lenses or on old observations will not be able to perform well on new observations, but the inclusion of DA will open doors for much easier discovery of new strong gravitational lenses.

Finally, with the increase in size and complexity of next-generation astronomical datasets, true comparison and understanding of these datasets will become increasingly difficult. The power of AI algorithms is in their ability to work and extract information from very complex multidimensional datasets. Our future work will focus on further development and refinement of DA algorithms. With their ability to extract similarities and differences between datasets, they can potentially be used to help us improve our simulations, and even understand the underlining physics.


\section*{Acknowledgments}

This manuscript has been supported by Fermi Research Alliance, LLC under Contract No.\ DE-AC02-07CH11359 with the U.S.\ Department of Energy (DOE), Office of Science, Office of High Energy Physics. This research has been partially supported by the High Velocity Artificial Intelligence grant as part of the DOE High Energy Physics Computational HEP program. 
This research has been partially supported by the DOE Office of Science, Office of Advanced Scientific Computing Research, applied mathematics and SciDAC programs under Contract No.\ DE-AC02-06CH11357. 
This research used resources of the Argonne Leadership Computing Facility at Argonne National Laboratory, which is a user facility supported by the DOE Office of Science.

The authors of this paper have committed themselves to performing this work in an equitable, inclusive, and just environment, and we hold ourselves accountable, believing that the best science is contingent on a good research environment.
We acknowledge the Deep Skies Lab as a community of multi-domain experts and collaborators who have facilitated an environment of open discussion, idea-generation, and collaboration. This community was important for the development of this project.

Furthermore, we also thank the two anonymous referees who helped improve this manuscript.

\subsection*{\it{Author Contributions}}

A.~\'Ciprijanovi\'c: \textit{Conceptualization, Data curation, Formal analysis, Investigation, Methodology, Project administration, Resources, Supervision, Visualization, Writing of original draft}; A.~Lewis: \textit{Formal analysis, Investigation, Methodology, Resources, Software, Visualization, Writing of original draft}; K.~Pedro: \textit{Conceptualization, Methodology, Project administration, Resources, Supervision, Writing  (review \& editing)}; S.~Madireddy: \textit{Conceptualization, Methodology, Resources, Supervision, Writing (review \& editing)}; B.~Nord: \textit{Conceptualization, Methodology, Supervision, Writing (review \& editing)}; G.~N.~Perdue: \textit{Conceptualization, Methodology, Project administration, Resources, Supervision, Writing  (review \& editing)}; S.~M.~Wild: \textit{Conceptualization, Methodology, Resources, Supervision, Writing (review \& editing)}.


\bibliographystyle{model2-names}
\bibliography{main} 

\begin{thebibliography}{89}
\expandafter\ifx\csname natexlab\endcsname\relax\def\natexlab#1{#1}\fi
\providecommand{\url}[1]{\texttt{#1}}
\providecommand{\href}[2]{#2}
\providecommand{\path}[1]{#1}
\providecommand{\DOIprefix}{doi:}
\providecommand{\ArXivprefix}{arXiv:}
\providecommand{\URLprefix}{URL: }
\providecommand{\Pubmedprefix}{pmid:}
\providecommand{\doi}[1]{\href{http://dx.doi.org/#1}{\path{#1}}}
\providecommand{\Pubmed}[1]{\href{pmid:#1}{\path{#1}}}
\providecommand{\bibinfo}[2]{#2}
\ifx\xfnm\relax \def\xfnm[#1]{\unskip,\space#1}\fi
\bibitem[{{Aihara} et~al.(2018){Aihara}, {Arimoto}, {Armstrong}, {Arnouts},
  {Bahcall}, {Bickerton}, {Bosch}, {Bundy} and et~al.}]{HSC2018}
\bibinfo{author}{{Aihara}, H.}, \bibinfo{author}{{Arimoto}, N.},
  \bibinfo{author}{{Armstrong}, R.}, \bibinfo{author}{{Arnouts}, S.},
  \bibinfo{author}{{Bahcall}, N.A.}, \bibinfo{author}{{Bickerton}, S.},
  \bibinfo{author}{{Bosch}, J.}, \bibinfo{author}{{Bundy}, K.},
  \bibinfo{author}{et~al.}, \bibinfo{year}{2018}.
\newblock \bibinfo{title}{{The Hyper Suprime-Cam SSP Survey: Overview and
  survey design}}.
\newblock \bibinfo{journal}{PASJ} \bibinfo{volume}{70}, \bibinfo{pages}{S4}.
\newblock \DOIprefix\doi{10.1093/pasj/psx066},
  \href{http://arxiv.org/abs/1704.05858}{ arXiv:1704.05858}.
\bibitem[{Aihara et~al.(2011)Aihara, Prieto, An, Anderson, Aubourg, Balbinot,
  Beers, Berlind, Bickerton, Bizyaev et~al.}]{AA2011}
\bibinfo{author}{Aihara, H.}, \bibinfo{author}{Prieto, C.A.},
  \bibinfo{author}{An, D.}, \bibinfo{author}{Anderson, S.F.},
  \bibinfo{author}{Aubourg, {\'{E}}.}, \bibinfo{author}{Balbinot, E.},
  \bibinfo{author}{Beers, T.C.}, \bibinfo{author}{Berlind, A.A.},
  \bibinfo{author}{Bickerton, S.J.}, \bibinfo{author}{Bizyaev, D.}, et~al.,
  \bibinfo{year}{2011}.
\newblock \bibinfo{title}{{Erratum}: The eight data release of the {Sloan}
  {Digital} {Sky} {Survey}: First data from {SDSS-III} (2011, {ApJS}, 193,
  29)}.
\newblock \bibinfo{journal}{The Astrophysical Journal Supplement Series}
  \bibinfo{volume}{195}, \bibinfo{pages}{26}.
\newblock \URLprefix \url{https://doi.org/10.1088/0067-0049/195/2/26},
  \DOIprefix\doi{10.1088/0067-0049/195/2/26}.
\bibitem[{{Alexander} et~al.(2021){Alexander}, {Gleyzer}, {Reddy}, {Tidball}
  and {Toomey}}]{AG2021}
\bibinfo{author}{{Alexander}, S.}, \bibinfo{author}{{Gleyzer}, S.},
  \bibinfo{author}{{Reddy}, P.}, \bibinfo{author}{{Tidball}, M.},
  \bibinfo{author}{{Toomey}, M.W.}, \bibinfo{year}{2021}.
\newblock \bibinfo{title}{{Domain Adaptation for Simulation-Based Dark Matter
  Searches Using Strong Gravitational Lensing}}.
\newblock \bibinfo{journal}{arXiv e-prints} ,
  \bibinfo{pages}{arXiv:2112.12121}\DOIprefix\doi{10.48550/arXiv.2112.12121},
  \href{http://arxiv.org/abs/2112.12121}{ arXiv:2112.12121}.
\bibitem[{Bergstra et~al.(2011)Bergstra, Bardenet, Bengio and
  K\'{e}gl}]{BB2011}
\bibinfo{author}{Bergstra, J.}, \bibinfo{author}{Bardenet, R.},
  \bibinfo{author}{Bengio, Y.}, \bibinfo{author}{K\'{e}gl, B.},
  \bibinfo{year}{2011}.
\newblock \bibinfo{title}{Algorithms for hyper-parameter optimization}, in:
  \bibinfo{editor}{Shawe-Taylor, J.}, \bibinfo{editor}{Zemel, R.},
  \bibinfo{editor}{Bartlett, P.}, \bibinfo{editor}{Pereira, F.},
  \bibinfo{editor}{Weinberger, K.} (Eds.), \bibinfo{booktitle}{Advances in
  Neural Information Processing Systems}, \bibinfo{publisher}{Curran
  Associates, Inc.}
\newblock \URLprefix
  \url{https://proceedings.neurips.cc/paper/2011/file/86e8f7ab32cfd12577bc2619bc635690-Paper.pdf}.
\bibitem[{{Bertin} and {Arnouts}(1996)}]{BA1996}
\bibinfo{author}{{Bertin}, E.}, \bibinfo{author}{{Arnouts}, S.},
  \bibinfo{year}{1996}.
\newblock \bibinfo{title}{{SExtractor: Software for source extraction.}}
\newblock \bibinfo{journal}{A\&AS} \bibinfo{volume}{117},
  \bibinfo{pages}{393--404}.
\newblock \DOIprefix\doi{10.1051/aas:1996164}.
\bibitem[{Borg and Groenen(2005)}]{BG2005}
\bibinfo{author}{Borg, I.}, \bibinfo{author}{Groenen, P.},
  \bibinfo{year}{2005}.
\newblock \bibinfo{title}{Modern Multidimensional Scaling: Theory and
  Applications (Springer Series in Statistics)}.
\newblock \DOIprefix\doi{10.1007/978-1-4757-2711-1}.
\bibitem[{{Busto} and {Gall}(2017)}]{BG2017}
\bibinfo{author}{{Busto}, P.P.}, \bibinfo{author}{{Gall}, J.},
  \bibinfo{year}{2017}.
\newblock \bibinfo{title}{Open set domain adaptation}, in:
  \bibinfo{booktitle}{2017 IEEE International Conference on Computer Vision
  (ICCV)}, pp. \bibinfo{pages}{754--763}.
\newblock \DOIprefix\doi{10.1109/ICCV.2017.88}.
\bibitem[{{Cao} et~al.(2019){Cao}, {You}, {Long}, {Wang} and {Yang}}]{CY2019}
\bibinfo{author}{{Cao}, Z.}, \bibinfo{author}{{You}, K.},
  \bibinfo{author}{{Long}, M.}, \bibinfo{author}{{Wang}, J.},
  \bibinfo{author}{{Yang}, Q.}, \bibinfo{year}{2019}.
\newblock \bibinfo{title}{{Learning to Transfer Examples for Partial Domain
  Adaptation}}.
\newblock \bibinfo{journal}{arXiv e-prints} ,
  \bibinfo{pages}{arXiv:1903.12230}\href{http://arxiv.org/abs/1903.12230}{
  arXiv:1903.12230}.
\bibitem[{{Cavanagh} et~al.(2021){Cavanagh}, {Bekki} and {Groves}}]{CB2021}
\bibinfo{author}{{Cavanagh}, M.K.}, \bibinfo{author}{{Bekki}, K.},
  \bibinfo{author}{{Groves}, B.A.}, \bibinfo{year}{2021}.
\newblock \bibinfo{title}{{Morphological classification of galaxies with deep
  learning: comparing 3-way and 4-way CNNs}}.
\newblock \bibinfo{journal}{MNRAS} \bibinfo{volume}{506},
  \bibinfo{pages}{659--676}.
\newblock \DOIprefix\doi{10.1093/mnras/stab1552},
  \href{http://arxiv.org/abs/2106.01571}{ arXiv:2106.01571}.
\bibitem[{{Cheng} et~al.(2021){Cheng}, {Conselice}, {Arag{\'o}n-Salamanca},
  {Aguena}, {Allam}, {Andrade-Oliveira}, {Annis}, {Bluck}, {Brooks}, {Burke}
  et~al.}]{CC2021}
\bibinfo{author}{{Cheng}, T.Y.}, \bibinfo{author}{{Conselice}, C.J.},
  \bibinfo{author}{{Arag{\'o}n-Salamanca}, A.}, \bibinfo{author}{{Aguena}, M.},
  \bibinfo{author}{{Allam}, S.}, \bibinfo{author}{{Andrade-Oliveira}, F.},
  \bibinfo{author}{{Annis}, J.}, \bibinfo{author}{{Bluck}, A.F.L.},
  \bibinfo{author}{{Brooks}, D.}, \bibinfo{author}{{Burke}, D.L.}, et~al.,
  \bibinfo{year}{2021}.
\newblock \bibinfo{title}{{Galaxy morphological classification catalogue of the
  Dark Energy Survey Year 3 data with convolutional neural networks}}.
\newblock \bibinfo{journal}{MNRAS} \bibinfo{volume}{507},
  \bibinfo{pages}{4425--4444}.
\newblock \DOIprefix\doi{10.1093/mnras/stab2142},
  \href{http://arxiv.org/abs/2107.10210}{ arXiv:2107.10210}.
\bibitem[{{{\'C}iprijanovi{\'c}} et~al.(2021){{\'C}iprijanovi{\'c}}, {Kafkes},
  {Downey}, {Jenkins}, {Perdue}, {Madireddy}, {Johnston}, {Snyder} and
  {Nord}}]{CK2021MNRAS}
\bibinfo{author}{{{\'C}iprijanovi{\'c}}, A.}, \bibinfo{author}{{Kafkes}, D.},
  \bibinfo{author}{{Downey}, K.}, \bibinfo{author}{{Jenkins}, S.},
  \bibinfo{author}{{Perdue}, G.N.}, \bibinfo{author}{{Madireddy}, S.},
  \bibinfo{author}{{Johnston}, T.}, \bibinfo{author}{{Snyder}, G.F.},
  \bibinfo{author}{{Nord}, B.}, \bibinfo{year}{2021}.
\newblock \bibinfo{title}{{DeepMerge - II. Building robust deep learning
  algorithms for merging galaxy identification across domains}}.
\newblock \bibinfo{journal}{MNRAS} \bibinfo{volume}{506},
  \bibinfo{pages}{677--691}.
\newblock \DOIprefix\doi{10.1093/mnras/stab1677},
  \href{http://arxiv.org/abs/2103.01373}{ arXiv:2103.01373}.
\bibitem[{{{\'C}iprijanovi{\'c}} et~al.(2022){{\'C}iprijanovi{\'c}}, {Kafkes},
  {Snyder}, {S{\'a}nchez}, {Perdue}, {Pedro}, {Nord}, {Madireddy} and
  {Wild}}]{CK2021}
\bibinfo{author}{{{\'C}iprijanovi{\'c}}, A.}, \bibinfo{author}{{Kafkes}, D.},
  \bibinfo{author}{{Snyder}, G.}, \bibinfo{author}{{S{\'a}nchez}, F.J.},
  \bibinfo{author}{{Perdue}, G.N.}, \bibinfo{author}{{Pedro}, K.},
  \bibinfo{author}{{Nord}, B.}, \bibinfo{author}{{Madireddy}, S.},
  \bibinfo{author}{{Wild}, S.M.}, \bibinfo{year}{2022}.
\newblock \bibinfo{title}{{DeepAdversaries: examining the robustness of deep
  learning models for galaxy morphology classification}}.
\newblock \bibinfo{journal}{Machine Learning: Science and Technology}
  \bibinfo{volume}{3}, \bibinfo{pages}{035007}.
\newblock \DOIprefix\doi{10.1088/2632-2153/ac7f1a},
  \href{http://arxiv.org/abs/2112.14299}{ arXiv:2112.14299}.
\bibitem[{{Conselice} et~al.(2003){Conselice}, {Bershady}, {Dickinson} and
  {Papovich}}]{CB2003}
\bibinfo{author}{{Conselice}, C.J.}, \bibinfo{author}{{Bershady}, M.A.},
  \bibinfo{author}{{Dickinson}, M.}, \bibinfo{author}{{Papovich}, C.},
  \bibinfo{year}{2003}.
\newblock \bibinfo{title}{{A Direct Measurement of Major Galaxy Mergers at z
  less than \~{}3}}.
\newblock \bibinfo{journal}{Astron. J.} \bibinfo{volume}{126},
  \bibinfo{pages}{1183--1207}.
\newblock \DOIprefix\doi{10.1086/377318},
  \href{http://arxiv.org/abs/astro-ph/0306106}{ arXiv:astro-ph/0306106}.
\bibitem[{Csurka(2017)}]{C2017}
\bibinfo{author}{Csurka, G.}, \bibinfo{year}{2017}.
\newblock \bibinfo{title}{A comprehensive survey on domain adaptation for
  visual applications}, in: \bibinfo{booktitle}{Domain Adaptation in Computer
  Vision Applications}. \bibinfo{publisher}{Springer International Publishing},
  \bibinfo{address}{Cham}, pp. \bibinfo{pages}{1--35}.
\newblock \DOIprefix\doi{10.1007/978-3-319-58347-1_1}.
\bibitem[{{Dark Energy Survey Collaboration} et~al.(2016){Dark Energy Survey
  Collaboration}, {Abbott}, {Abdalla}, {Aleksi{\'c}}, {Allam}, {Amara},
  {Bacon}, {Balbinot}, {Banerji} and et~al.}]{DES2016}
\bibinfo{author}{{Dark Energy Survey Collaboration}},
  \bibinfo{author}{{Abbott}, T.}, \bibinfo{author}{{Abdalla}, F.B.},
  \bibinfo{author}{{Aleksi{\'c}}, J.}, \bibinfo{author}{{Allam}, S.},
  \bibinfo{author}{{Amara}, A.}, \bibinfo{author}{{Bacon}, D.},
  \bibinfo{author}{{Balbinot}, E.}, \bibinfo{author}{{Banerji}, M.},
  \bibinfo{author}{et~al.}, \bibinfo{year}{2016}.
\newblock \bibinfo{title}{{The Dark Energy Survey: more than dark energy - an
  overview}}.
\newblock \bibinfo{journal}{MNRAS} \bibinfo{volume}{460},
  \bibinfo{pages}{1270--1299}.
\newblock \DOIprefix\doi{10.1093/MNRAS/stw641},
  \href{http://arxiv.org/abs/1601.00329}{ arXiv:1601.00329}.
\bibitem[{{Dav{\'e}} et~al.(2019){Dav{\'e}}, {Angl{\'e}s-Alc{\'a}zar},
  {Narayanan}, {Li}, {Rafieferantsoa} and {Appleby}}]{DA2019}
\bibinfo{author}{{Dav{\'e}}, R.}, \bibinfo{author}{{Angl{\'e}s-Alc{\'a}zar},
  D.}, \bibinfo{author}{{Narayanan}, D.}, \bibinfo{author}{{Li}, Q.},
  \bibinfo{author}{{Rafieferantsoa}, M.H.}, \bibinfo{author}{{Appleby}, S.},
  \bibinfo{year}{2019}.
\newblock \bibinfo{title}{{SIMBA: Cosmological simulations with black hole
  growth and feedback}}.
\newblock \bibinfo{journal}{MNRAS} \bibinfo{volume}{486},
  \bibinfo{pages}{2827--2849}.
\newblock \DOIprefix\doi{10.1093/mnras/stz937},
  \href{http://arxiv.org/abs/1901.10203}{ arXiv:1901.10203}.
\bibitem[{{Dey} et~al.(2019){Dey}, {Schlegel}, {Lang}, {Blum}, {Burleigh},
  {Fan}, {Findlay}, {Finkbeiner}, {Herrera}, {Juneau} et~al.}]{DS2019}
\bibinfo{author}{{Dey}, A.}, \bibinfo{author}{{Schlegel}, D.J.},
  \bibinfo{author}{{Lang}, D.}, \bibinfo{author}{{Blum}, R.},
  \bibinfo{author}{{Burleigh}, K.}, \bibinfo{author}{{Fan}, X.},
  \bibinfo{author}{{Findlay}, J.R.}, \bibinfo{author}{{Finkbeiner}, D.},
  \bibinfo{author}{{Herrera}, D.}, \bibinfo{author}{{Juneau}, S.}, et~al.,
  \bibinfo{year}{2019}.
\newblock \bibinfo{title}{{Overview of the DESI Legacy Imaging Surveys}}.
\newblock \bibinfo{journal}{Astron. J.} \bibinfo{volume}{157},
  \bibinfo{pages}{168}.
\newblock \DOIprefix\doi{10.3847/1538-3881/ab089d},
  \href{http://arxiv.org/abs/1804.08657}{ arXiv:1804.08657}.
\bibitem[{{Dickinson} et~al.(2018){Dickinson}, {Fortson}, {Lintott},
  {Scarlata}, {Willett}, {Bamford}, {Beck}, {Cardamone}, {Galloway}, {Simmons},
  {Keel}, {Kruk}, {Masters}, {Vogelsberger}, {Torrey} and {Snyder}}]{DF2018}
\bibinfo{author}{{Dickinson}, H.}, \bibinfo{author}{{Fortson}, L.},
  \bibinfo{author}{{Lintott}, C.}, \bibinfo{author}{{Scarlata}, C.},
  \bibinfo{author}{{Willett}, K.}, \bibinfo{author}{{Bamford}, S.},
  \bibinfo{author}{{Beck}, M.}, \bibinfo{author}{{Cardamone}, C.},
  \bibinfo{author}{{Galloway}, M.}, \bibinfo{author}{{Simmons}, B.},
  \bibinfo{author}{{Keel}, W.}, \bibinfo{author}{{Kruk}, S.},
  \bibinfo{author}{{Masters}, K.}, \bibinfo{author}{{Vogelsberger}, M.},
  \bibinfo{author}{{Torrey}, P.}, \bibinfo{author}{{Snyder}, G.F.},
  \bibinfo{year}{2018}.
\newblock \bibinfo{title}{{Galaxy Zoo: Morphological Classification of Galaxy
  Images from the Illustris Simulation}}.
\newblock \bibinfo{journal}{Astrophys. J.} \bibinfo{volume}{853},
  \bibinfo{pages}{194}.
\newblock \DOIprefix\doi{10.3847/1538-4357/aaa250},
  \href{http://arxiv.org/abs/1801.08541}{ arXiv:1801.08541}.
\bibitem[{{Dodge} and {Karam}(2016)}]{DK2016}
\bibinfo{author}{{Dodge}, S.}, \bibinfo{author}{{Karam}, L.},
  \bibinfo{year}{2016}.
\newblock \bibinfo{title}{{Understanding How Image Quality Affects Deep Neural
  Networks}}.
\newblock \bibinfo{journal}{arXiv e-prints}
  \href{http://arxiv.org/abs/1604.04004}{ arXiv:1604.04004}.
\bibitem[{{Dodge} and {Karam}(2017)}]{DK2017}
\bibinfo{author}{{Dodge}, S.}, \bibinfo{author}{{Karam}, L.},
  \bibinfo{year}{2017}.
\newblock \bibinfo{title}{A study and comparison of human and deep learning
  recognition performance under visual distortions}.
\newblock \bibinfo{journal}{arXiv e-prints}
  \href{http://arxiv.org/abs/1705.02498}{ arXiv:1705.02498}.
\bibitem[{{Ford} et~al.(2019){Ford}, {Gilmer}, {Carlini} and {Cubuk}}]{FG2019}
\bibinfo{author}{{Ford}, N.}, \bibinfo{author}{{Gilmer}, J.},
  \bibinfo{author}{{Carlini}, N.}, \bibinfo{author}{{Cubuk}, D.},
  \bibinfo{year}{2019}.
\newblock \bibinfo{title}{Adversarial examples are a natural consequence of
  test error in noise}.
\newblock \bibinfo{journal}{arXiv e-prints}
  \href{http://arxiv.org/abs/1901.10513}{ arXiv:1901.10513}.
\bibitem[{{Fu} et~al.(2019){Fu}, {Wu}, {Zhang} and {Yan}}]{FW2019}
\bibinfo{author}{{Fu}, J.}, \bibinfo{author}{{Wu}, X.},
  \bibinfo{author}{{Zhang}, S.}, \bibinfo{author}{{Yan}, J.},
  \bibinfo{year}{2019}.
\newblock \bibinfo{title}{Improved open set domain adaptation with
  backpropagation}, in: \bibinfo{booktitle}{2019 IEEE International Conference
  on Image Processing (ICIP)}, pp. \bibinfo{pages}{2506--2510}.
\newblock \DOIprefix\doi{10.1109/ICIP.2019.8803287}.
\bibitem[{{Galloway}(2017)}]{G2017}
\bibinfo{author}{{Galloway}, M.A.}, \bibinfo{year}{2017}.
\newblock \bibinfo{title}{{Morphology Is a Link to the Past: Examining
  Formative and Secular Galactic Evolution through Morphology}}.
\newblock Ph.D. thesis. University of Minnesota, Twin Cities.
\bibitem[{Ganin et~al.(2016)Ganin, Ustinova, Ajakan, Germain, Larochelle,
  Laviolette, March and Lempitsky}]{GU2016}
\bibinfo{author}{Ganin, Y.}, \bibinfo{author}{Ustinova, E.},
  \bibinfo{author}{Ajakan, H.}, \bibinfo{author}{Germain, P.},
  \bibinfo{author}{Larochelle, H.}, \bibinfo{author}{Laviolette, F.},
  \bibinfo{author}{March, M.}, \bibinfo{author}{Lempitsky, V.},
  \bibinfo{year}{2016}.
\newblock \bibinfo{title}{Domain-adversarial training of neural networks}.
\newblock \bibinfo{journal}{Journal of Machine Learning Research}
  \bibinfo{volume}{17}, \bibinfo{pages}{1--35}.
\newblock \URLprefix \url{http://jmlr.org/papers/v17/15-239.html}.
\bibitem[{{Gide} et~al.(2016){Gide}, {Dodge} and {Karam}}]{GD2016}
\bibinfo{author}{{Gide}, M.S.}, \bibinfo{author}{{Dodge}, S.F.},
  \bibinfo{author}{{Karam}, L.J.}, \bibinfo{year}{2016}.
\newblock \bibinfo{title}{The effect of distortions on the prediction of visual
  attention}.
\newblock \bibinfo{journal}{arXiv e-prints}
  \href{http://arxiv.org/abs/1604.03882}{ arXiv:1604.03882}.
\bibitem[{{Gilda} et~al.(2021){Gilda}, {de Mathelin}, {Bellstedt} and
  {Richard}}]{GB2021}
\bibinfo{author}{{Gilda}, S.}, \bibinfo{author}{{de Mathelin}, A.},
  \bibinfo{author}{{Bellstedt}, S.}, \bibinfo{author}{{Richard}, G.},
  \bibinfo{year}{2021}.
\newblock \bibinfo{title}{{Unsupervised Domain Adaptation for Constraining Star
  Formation Histories}}.
\newblock \bibinfo{journal}{arXiv e-prints} ,
  \bibinfo{pages}{arXiv:2112.14072}\DOIprefix\doi{10.48550/arXiv.2112.14072},
  \href{http://arxiv.org/abs/2112.14072}{ arXiv:2112.14072}.
\bibitem[{Gretton et~al.(2007)Gretton, Borgwardt, Rasch, Sch{\"o}lkopf and
  Smola}]{GB2007}
\bibinfo{author}{Gretton, A.}, \bibinfo{author}{Borgwardt, K.},
  \bibinfo{author}{Rasch, M.}, \bibinfo{author}{Sch{\"o}lkopf, B.},
  \bibinfo{author}{Smola, A.}, \bibinfo{year}{2007}.
\newblock \bibinfo{title}{A kernel method for the two-sample-problem}, in:
  \bibinfo{booktitle}{Advances in Neural Information Processing Systems 19},
  \bibinfo{publisher}{MIT Press}. pp. \bibinfo{pages}{513--520}.
\bibitem[{{Gretton} et~al.(2012){Gretton}, {Borgwardt}, {Rasch},
  {Sch{{\"o}}lkopf} and {Smola}}]{GB2008}
\bibinfo{author}{{Gretton}, A.}, \bibinfo{author}{{Borgwardt}, K.M.},
  \bibinfo{author}{{Rasch}, M.J.}, \bibinfo{author}{{Sch{{\"o}}lkopf}, B.},
  \bibinfo{author}{{Smola}, A.}, \bibinfo{year}{2012}.
\newblock \bibinfo{title}{A kernel two-sample test}.
\newblock \bibinfo{journal}{Journal of Machine Learning Research}
  \bibinfo{volume}{13}, \bibinfo{pages}{723--773}.
\newblock \URLprefix \url{http://jmlr.org/papers/v13/gretton12a.html}.
\bibitem[{{Han} et~al.(2020){Han}, {Rebuffi}, {Ehrhardt}, {Vedaldi} and
  {Zisserman}}]{HR2020}
\bibinfo{author}{{Han}, K.}, \bibinfo{author}{{Rebuffi}, S.A.},
  \bibinfo{author}{{Ehrhardt}, S.}, \bibinfo{author}{{Vedaldi}, A.},
  \bibinfo{author}{{Zisserman}, A.}, \bibinfo{year}{2020}.
\newblock \bibinfo{title}{{Automatically Discovering and Learning New Visual
  Categories with Ranking Statistics}}.
\newblock \bibinfo{journal}{arXiv e-prints} ,
  \bibinfo{pages}{arXiv:2002.05714}\href{http://arxiv.org/abs/2002.05714}{
  arXiv:2002.05714}.
\bibitem[{He et~al.(2016)He, Zhang, Ren and Sun}]{HZ2016}
\bibinfo{author}{He, K.}, \bibinfo{author}{Zhang, X.}, \bibinfo{author}{Ren,
  S.}, \bibinfo{author}{Sun, J.}, \bibinfo{year}{2016}.
\newblock \bibinfo{title}{Deep residual learning for image recognition}, in:
  \bibinfo{booktitle}{2016 IEEE Conference on Computer Vision and Pattern
  Recognition (CVPR)}, pp. \bibinfo{pages}{770--778}.
\newblock \DOIprefix\doi{10.1109/CVPR.2016.90}.
\bibitem[{{Holwerda} et~al.(2019){Holwerda}, {Kelvin}, {Baldry}, {Lintott},
  {Alpaslan}, {Pimbblet}, {Liske}, {Kitching}, {Bamford}, {de Jong}
  et~al.}]{HB2019A}
\bibinfo{author}{{Holwerda}, B.W.}, \bibinfo{author}{{Kelvin}, L.},
  \bibinfo{author}{{Baldry}, I.}, \bibinfo{author}{{Lintott}, C.},
  \bibinfo{author}{{Alpaslan}, M.}, \bibinfo{author}{{Pimbblet}, K.A.},
  \bibinfo{author}{{Liske}, J.}, \bibinfo{author}{{Kitching}, T.},
  \bibinfo{author}{{Bamford}, S.}, \bibinfo{author}{{de Jong}, J.}, et~al.,
  \bibinfo{year}{2019}.
\newblock \bibinfo{title}{{The Frequency of Dust Lanes in Edge-on Spiral
  Galaxies Identified by Galaxy Zoo in KiDS Imaging of GAMA Targets}}.
\newblock \bibinfo{journal}{Astron. J.} \bibinfo{volume}{158},
  \bibinfo{pages}{103}.
\newblock \DOIprefix\doi{10.3847/1538-3881/ab2886},
  \href{http://arxiv.org/abs/1909.07461}{ arXiv:1909.07461}.
\bibitem[{{Hubble}(1926)}]{H1926}
\bibinfo{author}{{Hubble}, E.P.}, \bibinfo{year}{1926}.
\newblock \bibinfo{title}{{Extragalactic nebulae.}}
\newblock \bibinfo{journal}{Astrophys. J.} \bibinfo{volume}{64},
  \bibinfo{pages}{321--369}.
\newblock \DOIprefix\doi{10.1086/143018}.
\bibitem[{{Huertas-Company} and {Lanusse}(2022)}]{HCL2022}
\bibinfo{author}{{Huertas-Company}, M.}, \bibinfo{author}{{Lanusse}, F.},
  \bibinfo{year}{2022}.
\newblock \bibinfo{title}{{The DAWES review 10: The impact of deep learning for
  the analysis of galaxy surveys}}.
\newblock \bibinfo{journal}{arXiv e-prints} ,
  \bibinfo{pages}{arXiv:2210.01813}\href{http://arxiv.org/abs/2210.01813}{
  arXiv:2210.01813}.
\bibitem[{{Ivezi{\'c}} et~al.(2019){Ivezi{\'c}}, {Kahn}, {Tyson}, {Abel},
  {Acosta}, {Allsman}, {Alonso}, {AlSayyad} and et~al.}]{IK2019}
\bibinfo{author}{{Ivezi{\'c}}, {\v{Z}}.}, \bibinfo{author}{{Kahn}, S.M.},
  \bibinfo{author}{{Tyson}, J.A.}, \bibinfo{author}{{Abel}, B.},
  \bibinfo{author}{{Acosta}, E.}, \bibinfo{author}{{Allsman}, R.},
  \bibinfo{author}{{Alonso}, D.}, \bibinfo{author}{{AlSayyad}, Y.},
  \bibinfo{author}{et~al.}, \bibinfo{year}{2019}.
\newblock \bibinfo{title}{{LSST}: From science drivers to reference design and
  anticipated data products}.
\newblock \bibinfo{journal}{Astrophys. J.} \bibinfo{volume}{873},
  \bibinfo{pages}{111}.
\newblock \DOIprefix\doi{10.3847/1538-4357/ab042c},
  \href{http://arxiv.org/abs/0805.2366}{ arXiv:0805.2366}.
\bibitem[{{Kauffmann} et~al.(2003){Kauffmann}, {Heckman}, {White}, {Charlot},
  {Tremonti}, {Peng}, {Seibert}, {Brinkmann}, {Nichol}, {SubbaRao} and
  {York}}]{KH2003}
\bibinfo{author}{{Kauffmann}, G.}, \bibinfo{author}{{Heckman}, T.M.},
  \bibinfo{author}{{White}, S.D.M.}, \bibinfo{author}{{Charlot}, S.},
  \bibinfo{author}{{Tremonti}, C.}, \bibinfo{author}{{Peng}, E.W.},
  \bibinfo{author}{{Seibert}, M.}, \bibinfo{author}{{Brinkmann}, J.},
  \bibinfo{author}{{Nichol}, R.C.}, \bibinfo{author}{{SubbaRao}, M.},
  \bibinfo{author}{{York}, D.}, \bibinfo{year}{2003}.
\newblock \bibinfo{title}{{The dependence of star formation history and
  internal structure on stellar mass for {}10$^{5}$ low-redshift galaxies}}.
\newblock \bibinfo{journal}{MNRAS} \bibinfo{volume}{341},
  \bibinfo{pages}{54--69}.
\newblock \DOIprefix\doi{10.1046/j.1365-8711.2003.06292.x},
  \href{http://arxiv.org/abs/astro-ph/0205070}{ arXiv:astro-ph/0205070}.
\bibitem[{Kullback and Leibler(1951)}]{KL1951}
\bibinfo{author}{Kullback, S.}, \bibinfo{author}{Leibler, R.A.},
  \bibinfo{year}{1951}.
\newblock \bibinfo{title}{On information and sufficiency}.
\newblock \bibinfo{journal}{Ann. Math. Statist.} \bibinfo{volume}{22},
  \bibinfo{pages}{79--86}.
\newblock \URLprefix \url{https://doi.org/10.1214/aoms/1177729694},
  \DOIprefix\doi{10.1214/aoms/1177729694}.
\bibitem[{{Li} et~al.(2021){Li}, {Li}, {Shi} and {Yu}}]{LL2021}
\bibinfo{author}{{Li}, J.}, \bibinfo{author}{{Li}, G.}, \bibinfo{author}{{Shi},
  Y.}, \bibinfo{author}{{Yu}, Y.}, \bibinfo{year}{2021}.
\newblock \bibinfo{title}{Cross-domain adaptive clustering for semi-supervised
  domain adaptation}, in: \bibinfo{booktitle}{Proceedings of the IEEE/CVF
  Conference on Computer Vision and Pattern Recognition (CVPR)}, pp.
  \bibinfo{pages}{2505--2514}.
\bibitem[{{Lianou} et~al.(2019){Lianou}, {Barmby}, {Mosenkov}, {Lehnert} and
  {Karczewski}}]{LB2019}
\bibinfo{author}{{Lianou}, S.}, \bibinfo{author}{{Barmby}, P.},
  \bibinfo{author}{{Mosenkov}, A.A.}, \bibinfo{author}{{Lehnert}, M.},
  \bibinfo{author}{{Karczewski}, O.}, \bibinfo{year}{2019}.
\newblock \bibinfo{title}{{Dust properties and star formation of approximately
  a thousand local galaxies}}.
\newblock \bibinfo{journal}{Astron. Astrophys.} \bibinfo{volume}{631},
  \bibinfo{pages}{A38}.
\newblock \DOIprefix\doi{10.1051/0004-6361/201834553},
  \href{http://arxiv.org/abs/1906.02712}{ arXiv:1906.02712}.
\bibitem[{Lintott et~al.(2010)Lintott, Schawinski, Bamford, Slosar, Land,
  Thomas, Edmondson, Masters, Nichol, Raddick, Szalay, Andreescu, Murray and
  Vandenberg}]{LS2010}
\bibinfo{author}{Lintott, C.}, \bibinfo{author}{Schawinski, K.},
  \bibinfo{author}{Bamford, S.}, \bibinfo{author}{Slosar, A.},
  \bibinfo{author}{Land, K.}, \bibinfo{author}{Thomas, D.},
  \bibinfo{author}{Edmondson, E.}, \bibinfo{author}{Masters, K.},
  \bibinfo{author}{Nichol, R.C.}, \bibinfo{author}{Raddick, M.J.},
  \bibinfo{author}{Szalay, A.}, \bibinfo{author}{Andreescu, D.},
  \bibinfo{author}{Murray, P.}, \bibinfo{author}{Vandenberg, J.},
  \bibinfo{year}{2010}.
\newblock \bibinfo{title}{{Galaxy Zoo 1: data release of morphological
  classifications for nearly 900 000 galaxies*}}.
\newblock \bibinfo{journal}{MNRAS} \bibinfo{volume}{410},
  \bibinfo{pages}{166--178}.
\newblock \URLprefix \url{https://doi.org/10.1111/j.1365-2966.2010.17432.x},
  \DOIprefix\doi{10.1111/j.1365-2966.2010.17432.x},
  \href{http://arxiv.org/abs/https://academic.oup.com/mnras/article-pdf/410/1/166/18442057/mnras0410-0166.pdf}{
  arXiv:https://academic.oup.com/mnras/article-pdf/410/1/166/18442057/mnras0410-0166.pdf}.
\bibitem[{{Lintott} et~al.(2011){Lintott}, {Schawinski}, {Bamford}, {Slosar},
  {Land}, {Thomas}, {Edmondson}, {Masters}, {Nichol}, {Raddick}, {Szalay},
  {Andreescu}, {Murray} and {Vandenberg}}]{LS2011}
\bibinfo{author}{{Lintott}, C.}, \bibinfo{author}{{Schawinski}, K.},
  \bibinfo{author}{{Bamford}, S.}, \bibinfo{author}{{Slosar}, A.},
  \bibinfo{author}{{Land}, K.}, \bibinfo{author}{{Thomas}, D.},
  \bibinfo{author}{{Edmondson}, E.}, \bibinfo{author}{{Masters}, K.},
  \bibinfo{author}{{Nichol}, R.C.}, \bibinfo{author}{{Raddick}, M.J.},
  \bibinfo{author}{{Szalay}, A.}, \bibinfo{author}{{Andreescu}, D.},
  \bibinfo{author}{{Murray}, P.}, \bibinfo{author}{{Vandenberg}, J.},
  \bibinfo{year}{2011}.
\newblock \bibinfo{title}{{Galaxy Zoo 1: data release of morphological
  classifications for nearly 900 000 galaxies}}.
\newblock \bibinfo{journal}{MNRAS} \bibinfo{volume}{410},
  \bibinfo{pages}{166--178}.
\newblock \DOIprefix\doi{10.1111/j.1365-2966.2010.17432.x},
  \href{http://arxiv.org/abs/1007.3265}{ arXiv:1007.3265}.
\bibitem[{{Lintott} et~al.(2008){Lintott}, {Schawinski}, {Slosar}, {Land},
  {Bamford}, {Thomas}, {Raddick}, {Nichol}, {Szalay}, {Andreescu}, {Murray} and
  {Vandenberg}}]{LS2008}
\bibinfo{author}{{Lintott}, C.J.}, \bibinfo{author}{{Schawinski}, K.},
  \bibinfo{author}{{Slosar}, A.}, \bibinfo{author}{{Land}, K.},
  \bibinfo{author}{{Bamford}, S.}, \bibinfo{author}{{Thomas}, D.},
  \bibinfo{author}{{Raddick}, M.J.}, \bibinfo{author}{{Nichol}, R.C.},
  \bibinfo{author}{{Szalay}, A.}, \bibinfo{author}{{Andreescu}, D.},
  \bibinfo{author}{{Murray}, P.}, \bibinfo{author}{{Vandenberg}, J.},
  \bibinfo{year}{2008}.
\newblock \bibinfo{title}{{Galaxy Zoo: morphologies derived from visual
  inspection of galaxies from the Sloan Digital Sky Survey}}.
\newblock \bibinfo{journal}{MNRAS} \bibinfo{volume}{389},
  \bibinfo{pages}{1179--1189}.
\newblock \DOIprefix\doi{10.1111/j.1365-2966.2008.13689.x},
  \href{http://arxiv.org/abs/0804.4483}{ arXiv:0804.4483}.
\bibitem[{{Liu} et~al.(2019){Liu}, {Cao}, {Long}, {Wang} and {Yang}}]{LC2019}
\bibinfo{author}{{Liu}, H.}, \bibinfo{author}{{Cao}, Z.},
  \bibinfo{author}{{Long}, M.}, \bibinfo{author}{{Wang}, J.},
  \bibinfo{author}{{Yang}, Q.}, \bibinfo{year}{2019}.
\newblock \bibinfo{title}{Separate to adapt: Open set domain adaptation via
  progressive separation}, in: \bibinfo{booktitle}{2019 IEEE/CVF Conference on
  Computer Vision and Pattern Recognition (CVPR)}, pp.
  \bibinfo{pages}{2922--2931}.
\newblock \DOIprefix\doi{10.1109/CVPR.2019.00304}.
\bibitem[{{Long} et~al.(2017){Long}, {Cao}, {Wang} and {Jordan}}]{LC2017}
\bibinfo{author}{{Long}, M.}, \bibinfo{author}{{Cao}, Z.},
  \bibinfo{author}{{Wang}, J.}, \bibinfo{author}{{Jordan}, M.I.},
  \bibinfo{year}{2017}.
\newblock \bibinfo{title}{Conditional adversarial domain adaptation}.
\newblock \bibinfo{journal}{arXiv e-prints}
  \href{http://arxiv.org/abs/1705.10667}{ arXiv:1705.10667}.
\bibitem[{{Lotz} et~al.(2004){Lotz}, {Primack} and {Madau}}]{LP2004}
\bibinfo{author}{{Lotz}, J.M.}, \bibinfo{author}{{Primack}, J.},
  \bibinfo{author}{{Madau}, P.}, \bibinfo{year}{2004}.
\newblock \bibinfo{title}{A new nonparametric approach to galaxy morphological
  classification}.
\newblock \bibinfo{journal}{Astron. J.} \bibinfo{volume}{128},
  \bibinfo{pages}{163--182}.
\newblock \DOIprefix\doi{10.1086/421849},
  \href{http://arxiv.org/abs/astro-ph/0311352}{ arXiv:astro-ph/0311352}.
\bibitem[{{Marinacci} et~al.(2018){Marinacci}, {Vogelsberger}, {Pakmor},
  {Torrey}, {Springel}, {Hernquist}, {Nelson}, {Weinberger} and
  et~al.}]{MV2018}
\bibinfo{author}{{Marinacci}, F.}, \bibinfo{author}{{Vogelsberger}, M.},
  \bibinfo{author}{{Pakmor}, R.}, \bibinfo{author}{{Torrey}, P.},
  \bibinfo{author}{{Springel}, V.}, \bibinfo{author}{{Hernquist}, L.},
  \bibinfo{author}{{Nelson}, D.}, \bibinfo{author}{{Weinberger}, R.},
  \bibinfo{author}{et~al.}, \bibinfo{year}{2018}.
\newblock \bibinfo{title}{{First results from the IllustrisTNG simulations:
  radio haloes and magnetic fields}}.
\newblock \bibinfo{journal}{MNRAS} \bibinfo{volume}{480},
  \bibinfo{pages}{5113--5139}.
\newblock \DOIprefix\doi{10.1093/MNRAS/sty2206},
  \href{http://arxiv.org/abs/1707.03396}{ arXiv:1707.03396}.
\bibitem[{{Naiman} et~al.(2018){Naiman}, {Pillepich}, {Springel},
  {Ramirez-Ruiz}, {Torrey}, {Vogelsberger}, {Pakmor}, {Nelson} and
  et~al.}]{NP2018}
\bibinfo{author}{{Naiman}, J.P.}, \bibinfo{author}{{Pillepich}, A.},
  \bibinfo{author}{{Springel}, V.}, \bibinfo{author}{{Ramirez-Ruiz}, E.},
  \bibinfo{author}{{Torrey}, P.}, \bibinfo{author}{{Vogelsberger}, M.},
  \bibinfo{author}{{Pakmor}, R.}, \bibinfo{author}{{Nelson}, D.},
  \bibinfo{author}{et~al.}, \bibinfo{year}{2018}.
\newblock \bibinfo{title}{{First results from the IllustrisTNG simulations: a
  tale of two elements - chemical evolution of magnesium and europium}}.
\newblock \bibinfo{journal}{MNRAS} \bibinfo{volume}{477},
  \bibinfo{pages}{1206--1224}.
\newblock \DOIprefix\doi{10.1093/MNRAS/sty618},
  \href{http://arxiv.org/abs/1707.03401}{ arXiv:1707.03401}.
\bibitem[{{Nelson} et~al.(2015){Nelson}, {Pillepich}, {Genel}, {Vogelsberger},
  {Springel}, {Torrey}, {Rodriguez-Gomez}, {Sijacki}, {Snyder}, {Griffen},
  {Marinacci}, {Blecha}, {Sales}, {Xu} and {Hernquist}}]{NP2015}
\bibinfo{author}{{Nelson}, D.}, \bibinfo{author}{{Pillepich}, A.},
  \bibinfo{author}{{Genel}, S.}, \bibinfo{author}{{Vogelsberger}, M.},
  \bibinfo{author}{{Springel}, V.}, \bibinfo{author}{{Torrey}, P.},
  \bibinfo{author}{{Rodriguez-Gomez}, V.}, \bibinfo{author}{{Sijacki}, D.},
  \bibinfo{author}{{Snyder}, G.F.}, \bibinfo{author}{{Griffen}, B.},
  \bibinfo{author}{{Marinacci}, F.}, \bibinfo{author}{{Blecha}, L.},
  \bibinfo{author}{{Sales}, L.}, \bibinfo{author}{{Xu}, D.},
  \bibinfo{author}{{Hernquist}, L.}, \bibinfo{year}{2015}.
\newblock \bibinfo{title}{{The illustris simulation: Public data release}}.
\newblock \bibinfo{journal}{Astronomy and Computing} \bibinfo{volume}{13},
  \bibinfo{pages}{12--37}.
\newblock \DOIprefix\doi{10.1016/j.ascom.2015.09.003},
  \href{http://arxiv.org/abs/1504.00362}{ arXiv:1504.00362}.
\bibitem[{{Nelson} et~al.(2019){Nelson}, {Springel}, {Pillepich},
  {Rodriguez-Gomez}, {Torrey}, {Genel}, {Vogelsberger}, {Pakmor}, {Marinacci},
  {Weinberger}, {Kelley}, {Lovell}, {Diemer} and {Hernquist}}]{NS2019}
\bibinfo{author}{{Nelson}, D.}, \bibinfo{author}{{Springel}, V.},
  \bibinfo{author}{{Pillepich}, A.}, \bibinfo{author}{{Rodriguez-Gomez}, V.},
  \bibinfo{author}{{Torrey}, P.}, \bibinfo{author}{{Genel}, S.},
  \bibinfo{author}{{Vogelsberger}, M.}, \bibinfo{author}{{Pakmor}, R.},
  \bibinfo{author}{{Marinacci}, F.}, \bibinfo{author}{{Weinberger}, R.},
  \bibinfo{author}{{Kelley}, L.}, \bibinfo{author}{{Lovell}, M.},
  \bibinfo{author}{{Diemer}, B.}, \bibinfo{author}{{Hernquist}, L.},
  \bibinfo{year}{2019}.
\newblock \bibinfo{title}{{The IllustrisTNG simulations: public data release}}.
\newblock \bibinfo{journal}{Computational Astrophysics and Cosmology}
  \bibinfo{volume}{6}, \bibinfo{pages}{2}.
\newblock \DOIprefix\doi{10.1186/s40668-019-0028-x},
  \href{http://arxiv.org/abs/1812.05609}{ arXiv:1812.05609}.
\bibitem[{Pedregosa et~al.(2011)Pedregosa, Varoquaux, Gramfort, Michel,
  Thirion, Grisel, Blondel, Prettenhofer and et~al.}]{PV2011}
\bibinfo{author}{Pedregosa, F.}, \bibinfo{author}{Varoquaux, G.},
  \bibinfo{author}{Gramfort, A.}, \bibinfo{author}{Michel, V.},
  \bibinfo{author}{Thirion, B.}, \bibinfo{author}{Grisel, O.},
  \bibinfo{author}{Blondel, M.}, \bibinfo{author}{Prettenhofer, P.},
  \bibinfo{author}{et~al.}, \bibinfo{year}{2011}.
\newblock \bibinfo{title}{Scikit-learn: Machine learning in {P}ython}.
\newblock \bibinfo{journal}{Journal of Machine Learning Research}
  \bibinfo{volume}{12}, \bibinfo{pages}{2825--2830}.
\bibitem[{{Peng} et~al.(2017){Peng}, {Usman}, {Kaushik}, {Hoffman}, {Wang} and
  {Saenko}}]{PU2017}
\bibinfo{author}{{Peng}, X.}, \bibinfo{author}{{Usman}, B.},
  \bibinfo{author}{{Kaushik}, N.}, \bibinfo{author}{{Hoffman}, J.},
  \bibinfo{author}{{Wang}, D.}, \bibinfo{author}{{Saenko}, K.},
  \bibinfo{year}{2017}.
\newblock \bibinfo{title}{{VisDA: The Visual Domain Adaptation Challenge}}.
\newblock \bibinfo{journal}{arXiv e-prints} ,
  \bibinfo{pages}{arXiv:1710.06924}\href{http://arxiv.org/abs/1710.06924}{
  arXiv:1710.06924}.
\bibitem[{{Pillepich} et~al.(2018){Pillepich}, {Nelson}, {Hernquist},
  {Springel}, {Pakmor}, {Torrey}, {Weinberger}, {Genel} and et~al.}]{PH2018}
\bibinfo{author}{{Pillepich}, A.}, \bibinfo{author}{{Nelson}, D.},
  \bibinfo{author}{{Hernquist}, L.}, \bibinfo{author}{{Springel}, V.},
  \bibinfo{author}{{Pakmor}, R.}, \bibinfo{author}{{Torrey}, P.},
  \bibinfo{author}{{Weinberger}, R.}, \bibinfo{author}{{Genel}, S.},
  \bibinfo{author}{et~al.}, \bibinfo{year}{2018}.
\newblock \bibinfo{title}{{First results from the IllustrisTNG simulations: the
  stellar mass content of groups and clusters of galaxies}}.
\newblock \bibinfo{journal}{MNRAS} \bibinfo{volume}{475},
  \bibinfo{pages}{648--675}.
\newblock \DOIprefix\doi{10.1093/MNRAS/stx3112},
  \href{http://arxiv.org/abs/1707.03406}{ arXiv:1707.03406}.
\bibitem[{{Rowe} et~al.(2015){Rowe}, {Jarvis}, {Mandelbaum}, {Bernstein},
  {Bosch}, {Simet}, {Meyers}, {Kacprzak} and et~al.}]{RJ2015}
\bibinfo{author}{{Rowe}, B.T.P.}, \bibinfo{author}{{Jarvis}, M.},
  \bibinfo{author}{{Mandelbaum}, R.}, \bibinfo{author}{{Bernstein}, G.M.},
  \bibinfo{author}{{Bosch}, J.}, \bibinfo{author}{{Simet}, M.},
  \bibinfo{author}{{Meyers}, J.E.}, \bibinfo{author}{{Kacprzak}, T.},
  \bibinfo{author}{et~al.}, \bibinfo{year}{2015}.
\newblock \bibinfo{title}{{GALSIM: The modular galaxy image simulation
  toolkit}}.
\newblock \bibinfo{journal}{Astronomy and Computing} \bibinfo{volume}{10},
  \bibinfo{pages}{121--150}.
\newblock \DOIprefix\doi{10.1016/j.ascom.2015.02.002},
  \href{http://arxiv.org/abs/1407.7676}{ arXiv:1407.7676}.
\bibitem[{{Saenko} et~al.(2010){Saenko}, {Kulis}, {Fritz} and
  {Darrell}}]{SK2010}
\bibinfo{author}{{Saenko}, K.}, \bibinfo{author}{{Kulis}, B.},
  \bibinfo{author}{{Fritz}, M.}, \bibinfo{author}{{Darrell}, T.},
  \bibinfo{year}{2010}.
\newblock \bibinfo{title}{Adapting visual category models to new domains}, in:
  \bibinfo{booktitle}{Proceedings of the 11th European Conference on Computer
  Vision: Part IV}, \bibinfo{publisher}{Springer-Verlag},
  \bibinfo{address}{Berlin, Heidelberg}. p. \bibinfo{pages}{213–226}.
\bibitem[{{Saito} et~al.(2020){Saito}, {Kim}, {Sclaroff} and {Saenko}}]{SK2020}
\bibinfo{author}{{Saito}, K.}, \bibinfo{author}{{Kim}, D.},
  \bibinfo{author}{{Sclaroff}, S.}, \bibinfo{author}{{Saenko}, K.},
  \bibinfo{year}{2020}.
\newblock \bibinfo{title}{{Universal Domain Adaptation through Self
  Supervision}}.
\newblock \bibinfo{journal}{arXiv e-prints} ,
  \bibinfo{pages}{arXiv:2002.07953}\href{http://arxiv.org/abs/2002.07953}{
  arXiv:2002.07953}.
\bibitem[{{Schaye} et~al.(2015){Schaye}, {Crain}, {Bower}, {Furlong},
  {Schaller}, {Theuns}, {Dalla Vecchia}, {Frenk}, {McCarthy}, {Helly}
  et~al.}]{SC2015}
\bibinfo{author}{{Schaye}, J.}, \bibinfo{author}{{Crain}, R.A.},
  \bibinfo{author}{{Bower}, R.G.}, \bibinfo{author}{{Furlong}, M.},
  \bibinfo{author}{{Schaller}, M.}, \bibinfo{author}{{Theuns}, T.},
  \bibinfo{author}{{Dalla Vecchia}, C.}, \bibinfo{author}{{Frenk}, C.S.},
  \bibinfo{author}{{McCarthy}, I.G.}, \bibinfo{author}{{Helly}, J.C.}, et~al.,
  \bibinfo{year}{2015}.
\newblock \bibinfo{title}{{The EAGLE project: simulating the evolution and
  assembly of galaxies and their environments}}.
\newblock \bibinfo{journal}{MNRAS} \bibinfo{volume}{446},
  \bibinfo{pages}{521--554}.
\newblock \DOIprefix\doi{10.1093/mnras/stu2058},
  \href{http://arxiv.org/abs/1407.7040}{ arXiv:1407.7040}.
\bibitem[{{S{\'e}rsic}(1963)}]{S1963}
\bibinfo{author}{{S{\'e}rsic}, J.L.}, \bibinfo{year}{1963}.
\newblock \bibinfo{title}{{Photometry of southern galaxies IX: NGC 1313}}.
\newblock \bibinfo{journal}{Boletin de la Asociacion Argentina de Astronomia La
  Plata Argentina} \bibinfo{volume}{6}, \bibinfo{pages}{99--99}.
\bibitem[{Shahriari et~al.(2016)Shahriari, Swersky, Wang, Adams and
  de~Freitas}]{SS2016bayes}
\bibinfo{author}{Shahriari, B.}, \bibinfo{author}{Swersky, K.},
  \bibinfo{author}{Wang, Z.}, \bibinfo{author}{Adams, R.P.},
  \bibinfo{author}{de~Freitas, N.}, \bibinfo{year}{2016}.
\newblock \bibinfo{title}{Taking the human out of the loop: A review of
  bayesian optimization}.
\newblock \bibinfo{journal}{Proceedings of the IEEE} \bibinfo{volume}{104},
  \bibinfo{pages}{148--175}.
\newblock \DOIprefix\doi{10.1109/JPROC.2015.2494218}.
\bibitem[{{Simmons} et~al.(2017){Simmons}, {Lintott}, {Willett}, {Masters},
  {Kartaltepe}, {H{\"a}u{\ss}ler}, {Kaviraj}, {Krawczyk}, {Kruk}, {McIntosh}
  et~al.}]{SL2017}
\bibinfo{author}{{Simmons}, B.D.}, \bibinfo{author}{{Lintott}, C.},
  \bibinfo{author}{{Willett}, K.W.}, \bibinfo{author}{{Masters}, K.L.},
  \bibinfo{author}{{Kartaltepe}, J.S.}, \bibinfo{author}{{H{\"a}u{\ss}ler},
  B.}, \bibinfo{author}{{Kaviraj}, S.}, \bibinfo{author}{{Krawczyk}, C.},
  \bibinfo{author}{{Kruk}, S.J.}, \bibinfo{author}{{McIntosh}, D.H.}, et~al.,
  \bibinfo{year}{2017}.
\newblock \bibinfo{title}{{Galaxy Zoo: quantitative visual morphological
  classifications for 48 000 galaxies from CANDELS}}.
\newblock \bibinfo{journal}{MNRAS} \bibinfo{volume}{464},
  \bibinfo{pages}{4420--4447}.
\newblock \DOIprefix\doi{10.1093/mnras/stw2587},
  \href{http://arxiv.org/abs/1610.03070}{ arXiv:1610.03070}.
\bibitem[{{Slijepcevic} et~al.(2022){Slijepcevic}, {Scaife}, {Walmsley} and
  {Bowles}}]{SS2022}
\bibinfo{author}{{Slijepcevic}, I.V.}, \bibinfo{author}{{Scaife}, A.M.M.},
  \bibinfo{author}{{Walmsley}, M.}, \bibinfo{author}{{Bowles}, M.},
  \bibinfo{year}{2022}.
\newblock \bibinfo{title}{{Learning useful representations for radio astronomy
  ``in the wild'' with contrastive learning}}.
\newblock \bibinfo{journal}{arXiv e-prints} ,
  \bibinfo{pages}{arXiv:2207.08666}\DOIprefix\doi{10.48550/arXiv.2207.08666},
  \href{http://arxiv.org/abs/2207.08666}{ arXiv:2207.08666}.
\bibitem[{{Snyder} et~al.(2019){Snyder}, {Rodriguez-Gomez}, {Lotz}, {Torrey},
  {Quirk}, {Hernquist}, {Vogelsberger} and {Freeman}}]{SR2019}
\bibinfo{author}{{Snyder}, G.F.}, \bibinfo{author}{{Rodriguez-Gomez}, V.},
  \bibinfo{author}{{Lotz}, J.M.}, \bibinfo{author}{{Torrey}, P.},
  \bibinfo{author}{{Quirk}, A.C.N.}, \bibinfo{author}{{Hernquist}, L.},
  \bibinfo{author}{{Vogelsberger}, M.}, \bibinfo{author}{{Freeman}, P.E.},
  \bibinfo{year}{2019}.
\newblock \bibinfo{title}{{Automated distant galaxy merger classifications from
  Space Telescope images using the Illustris simulation}}.
\newblock \bibinfo{journal}{MNRAS} \bibinfo{volume}{486},
  \bibinfo{pages}{3702--3720}.
\newblock \DOIprefix\doi{10.1093/mnras/stz1059},
  \href{http://arxiv.org/abs/1809.02136}{ arXiv:1809.02136}.
\bibitem[{Snyder et~al.(2015)Snyder, Torrey, Lotz, Genel, McBride,
  Vogelsberger, Pillepich, Nelson and et~al.}]{ST2015}
\bibinfo{author}{Snyder, G.F.}, \bibinfo{author}{Torrey, P.},
  \bibinfo{author}{Lotz, J.M.}, \bibinfo{author}{Genel, S.},
  \bibinfo{author}{McBride, C.K.}, \bibinfo{author}{Vogelsberger, M.},
  \bibinfo{author}{Pillepich, A.}, \bibinfo{author}{Nelson, D.},
  \bibinfo{author}{et~al.}, \bibinfo{year}{2015}.
\newblock \bibinfo{title}{{Galaxy morphology and star formation in the
  Illustris Simulation at z = 0}}.
\newblock \bibinfo{journal}{MNRAS} \bibinfo{volume}{454},
  \bibinfo{pages}{1886--1908}.
\newblock \DOIprefix\doi{10.1093/MNRAS/stv2078}.
\bibitem[{{Springel}(2010)}]{S2010}
\bibinfo{author}{{Springel}, V.}, \bibinfo{year}{2010}.
\newblock \bibinfo{title}{{E pur si muove: Galilean-invariant cosmological
  hydrodynamical simulations on a moving mesh}}.
\newblock \bibinfo{journal}{MNRAS} \bibinfo{volume}{401},
  \bibinfo{pages}{791--851}.
\newblock \DOIprefix\doi{10.1111/j.1365-2966.2009.15715.x},
  \href{http://arxiv.org/abs/0901.4107}{ arXiv:0901.4107}.
\bibitem[{{Springel} et~al.(2018){Springel}, {Pakmor}, {Pillepich},
  {Weinberger}, {Nelson}, {Hernquist}, {Vogelsberger}, {Genel} and
  et~al.}]{SP2018}
\bibinfo{author}{{Springel}, V.}, \bibinfo{author}{{Pakmor}, R.},
  \bibinfo{author}{{Pillepich}, A.}, \bibinfo{author}{{Weinberger}, R.},
  \bibinfo{author}{{Nelson}, D.}, \bibinfo{author}{{Hernquist}, L.},
  \bibinfo{author}{{Vogelsberger}, M.}, \bibinfo{author}{{Genel}, S.},
  \bibinfo{author}{et~al.}, \bibinfo{year}{2018}.
\newblock \bibinfo{title}{{First results from the IllustrisTNG simulations:
  matter and galaxy clustering}}.
\newblock \bibinfo{journal}{MNRAS} \bibinfo{volume}{475},
  \bibinfo{pages}{676--698}.
\newblock \DOIprefix\doi{10.1093/MNRAS/stx3304},
  \href{http://arxiv.org/abs/1707.03397}{ arXiv:1707.03397}.
\bibitem[{Sugiyama et~al.(2007)Sugiyama, Nakajima, Kashima, Buenau and
  Kawanabe}]{SN2007}
\bibinfo{author}{Sugiyama, M.}, \bibinfo{author}{Nakajima, S.},
  \bibinfo{author}{Kashima, H.}, \bibinfo{author}{Buenau, P.},
  \bibinfo{author}{Kawanabe, M.}, \bibinfo{year}{2007}.
\newblock \bibinfo{title}{Direct importance estimation with model selection and
  its application to covariate shift adaptation}, in: \bibinfo{editor}{Platt,
  J.}, \bibinfo{editor}{Koller, D.}, \bibinfo{editor}{Singer, Y.},
  \bibinfo{editor}{Roweis, S.} (Eds.), \bibinfo{booktitle}{Advances in Neural
  Information Processing Systems}, \bibinfo{publisher}{Curran Associates, Inc.}
\newblock \URLprefix
  \url{https://proceedings.neurips.cc/paper/2007/file/be83ab3ecd0db773eb2dc1b0a17836a1-Paper.pdf}.
\bibitem[{Sun and Saenko(2016)}]{SS2016}
\bibinfo{author}{Sun, B.}, \bibinfo{author}{Saenko, K.}, \bibinfo{year}{2016}.
\newblock \bibinfo{title}{Deep {CORAL}: Correlation alignment for deep domain
  adaptation}, in: \bibinfo{booktitle}{ECCV Workshops}.
\bibitem[{Sutskever et~al.(2013)Sutskever, Martens, Dahl and Hinton}]{SM2013}
\bibinfo{author}{Sutskever, I.}, \bibinfo{author}{Martens, J.},
  \bibinfo{author}{Dahl, G.}, \bibinfo{author}{Hinton, G.},
  \bibinfo{year}{2013}.
\newblock \bibinfo{title}{On the importance of initialization and momentum in
  deep learning}, in: \bibinfo{editor}{Dasgupta, S.},
  \bibinfo{editor}{McAllester, D.} (Eds.), \bibinfo{booktitle}{Proceedings of
  the 30th International Conference on Machine Learning},
  \bibinfo{publisher}{PMLR}, \bibinfo{address}{Atlanta, Georgia, USA}. pp.
  \bibinfo{pages}{1139--1147}.
\newblock \URLprefix \url{https://proceedings.mlr.press/v28/sutskever13.html}.
\bibitem[{{Tenenbaum} et~al.(2000){Tenenbaum}, {de Silva} and
  {Langford}}]{TS2000}
\bibinfo{author}{{Tenenbaum}, J.B.}, \bibinfo{author}{{de Silva}, V.},
  \bibinfo{author}{{Langford}, J.C.}, \bibinfo{year}{2000}.
\newblock \bibinfo{title}{{A Global Geometric Framework for Nonlinear
  Dimensionality Reduction}}.
\newblock \bibinfo{journal}{Science} \bibinfo{volume}{290},
  \bibinfo{pages}{2319--2323}.
\newblock \DOIprefix\doi{10.1126/science.290.5500.2319}.
\bibitem[{{Thota} and {Leontidis}(2021)}]{TL2021}
\bibinfo{author}{{Thota}, M.}, \bibinfo{author}{{Leontidis}, G.},
  \bibinfo{year}{2021}.
\newblock \bibinfo{title}{{Contrastive Domain Adaptation}}.
\newblock \bibinfo{journal}{arXiv e-prints} ,
  \bibinfo{pages}{arXiv:2103.15566}\DOIprefix\doi{10.48550/arXiv.2103.15566},
  \href{http://arxiv.org/abs/2103.15566}{ arXiv:2103.15566}.
\bibitem[{{Tian} et~al.(2019){Tian}, {Krishnan} and {Isola}}]{TK2019}
\bibinfo{author}{{Tian}, Y.}, \bibinfo{author}{{Krishnan}, D.},
  \bibinfo{author}{{Isola}, P.}, \bibinfo{year}{2019}.
\newblock \bibinfo{title}{{Contrastive Multiview Coding}}.
\newblock \bibinfo{journal}{arXiv e-prints} ,
  \bibinfo{pages}{arXiv:1906.05849}\DOIprefix\doi{10.48550/arXiv.1906.05849},
  \href{http://arxiv.org/abs/1906.05849}{ arXiv:1906.05849}.
\bibitem[{{van den Bergh}(1960)}]{VB1960}
\bibinfo{author}{{van den Bergh}, S.}, \bibinfo{year}{1960}.
\newblock \bibinfo{title}{{A Preliminary Liminosity Classification for Galaxies
  of Type Sb.}}
\newblock \bibinfo{journal}{Astrophys. J.} \bibinfo{volume}{131},
  \bibinfo{pages}{558}.
\newblock \DOIprefix\doi{10.1086/146869}.
\bibitem[{{van der Maaten} and {Hinton}(2008)}]{MH2008}
\bibinfo{author}{{van der Maaten}, L.}, \bibinfo{author}{{Hinton}, G.},
  \bibinfo{year}{2008}.
\newblock \bibinfo{title}{Visualizing data using {t-SNE}}.
\newblock \bibinfo{journal}{Journal of Machine Learning Research}
  \bibinfo{volume}{9}, \bibinfo{pages}{2579--2605}.
\newblock \URLprefix \url{http://jmlr.org/papers/v9/vandermaaten08a.html}.
\bibitem[{{Venkateswara} et~al.(2017){Venkateswara}, {Eusebio}, {Chakraborty}
  and {Panchanathan}}]{VE2017}
\bibinfo{author}{{Venkateswara}, H.}, \bibinfo{author}{{Eusebio}, J.},
  \bibinfo{author}{{Chakraborty}, S.}, \bibinfo{author}{{Panchanathan}, S.},
  \bibinfo{year}{2017}.
\newblock \bibinfo{title}{Deep hashing network for unsupervised domain
  adaptation}.
\newblock \bibinfo{journal}{2017 IEEE Conference on Computer Vision and Pattern
  Recognition (CVPR)} , \bibinfo{pages}{5385--5394}.
\bibitem[{{Vilalta} et~al.(2019){Vilalta}, {Dhar Gupta}, {Boumber} and
  {Meskhi}}]{VD2019}
\bibinfo{author}{{Vilalta}, R.}, \bibinfo{author}{{Dhar Gupta}, K.},
  \bibinfo{author}{{Boumber}, D.}, \bibinfo{author}{{Meskhi}, M.M.},
  \bibinfo{year}{2019}.
\newblock \bibinfo{title}{{A General Approach to Domain Adaptation with
  Applications in Astronomy}}.
\newblock \bibinfo{journal}{PASP} \bibinfo{volume}{131},
  \bibinfo{pages}{108008}.
\newblock \DOIprefix\doi{10.1088/1538-3873/aaf1fc},
  \href{http://arxiv.org/abs/1812.08839}{ arXiv:1812.08839}.
\bibitem[{{Vogelsberger} et~al.(2014){Vogelsberger}, {Genel}, {Springel},
  {Torrey}, {Sijacki}, {Xu}, {Snyder}, {Nelson} and {Hernquist}}]{VG2014}
\bibinfo{author}{{Vogelsberger}, M.}, \bibinfo{author}{{Genel}, S.},
  \bibinfo{author}{{Springel}, V.}, \bibinfo{author}{{Torrey}, P.},
  \bibinfo{author}{{Sijacki}, D.}, \bibinfo{author}{{Xu}, D.},
  \bibinfo{author}{{Snyder}, G.}, \bibinfo{author}{{Nelson}, D.},
  \bibinfo{author}{{Hernquist}, L.}, \bibinfo{year}{2014}.
\newblock \bibinfo{title}{{Introducing the Illustris Project: simulating the
  coevolution of dark and visible matter in the Universe}}.
\newblock \bibinfo{journal}{MNRAS} \bibinfo{volume}{444},
  \bibinfo{pages}{1518--1547}.
\newblock \DOIprefix\doi{10.1093/mnras/stu1536},
  \href{http://arxiv.org/abs/1405.2921}{ arXiv:1405.2921}.
\bibitem[{{Walmsley} et~al.(2022a){Walmsley}, {Lintott}, {G{\'e}ron}, {Kruk},
  {Krawczyk}, {Willett}, {Bamford}, {Kelvin}, {Fortson}, {Gal}, {Keel},
  {Masters}, {Mehta}, {Simmons}, {Smethurst}, {Smith}, {Baeten} and
  {Macmillan}}]{WL2022}
\bibinfo{author}{{Walmsley}, M.}, \bibinfo{author}{{Lintott}, C.},
  \bibinfo{author}{{G{\'e}ron}, T.}, \bibinfo{author}{{Kruk}, S.},
  \bibinfo{author}{{Krawczyk}, C.}, \bibinfo{author}{{Willett}, K.W.},
  \bibinfo{author}{{Bamford}, S.}, \bibinfo{author}{{Kelvin}, L.S.},
  \bibinfo{author}{{Fortson}, L.}, \bibinfo{author}{{Gal}, Y.},
  \bibinfo{author}{{Keel}, W.}, \bibinfo{author}{{Masters}, K.L.},
  \bibinfo{author}{{Mehta}, V.}, \bibinfo{author}{{Simmons}, B.D.},
  \bibinfo{author}{{Smethurst}, R.}, \bibinfo{author}{{Smith}, L.},
  \bibinfo{author}{{Baeten}, E.M.}, \bibinfo{author}{{Macmillan}, C.},
  \bibinfo{year}{2022}a.
\newblock \bibinfo{title}{{Galaxy Zoo DECaLS: Detailed visual morphology
  measurements from volunteers and deep learning for 314 000 galaxies}}.
\newblock \bibinfo{journal}{MNRAS} \bibinfo{volume}{509},
  \bibinfo{pages}{3966--3988}.
\newblock \DOIprefix\doi{10.1093/mnras/stab2093},
  \href{http://arxiv.org/abs/2102.08414}{ arXiv:2102.08414}.
\bibitem[{{Walmsley} et~al.(2022b){Walmsley}, {Slijepcevic}, {Bowles} and
  {Scaife}}]{VS2022}
\bibinfo{author}{{Walmsley}, M.}, \bibinfo{author}{{Slijepcevic}, I.V.},
  \bibinfo{author}{{Bowles}, M.}, \bibinfo{author}{{Scaife}, A.M.M.},
  \bibinfo{year}{2022}b.
\newblock \bibinfo{title}{{Towards Galaxy Foundation Models with Hybrid
  Contrastive Learning}}.
\newblock \bibinfo{journal}{arXiv e-prints} ,
  \bibinfo{pages}{arXiv:2206.11927}\DOIprefix\doi{10.48550/arXiv.2206.11927},
  \href{http://arxiv.org/abs/2206.11927}{ arXiv:2206.11927}.
\bibitem[{{Walmsley} et~al.(2020){Walmsley}, {Smith}, {Lintott}, {Gal},
  {Bamford}, {Dickinson}, {Fortson}, {Kruk}, {Masters}, {Scarlata}, {Simmons},
  {Smethurst} and {Wright}}]{WS2020}
\bibinfo{author}{{Walmsley}, M.}, \bibinfo{author}{{Smith}, L.},
  \bibinfo{author}{{Lintott}, C.}, \bibinfo{author}{{Gal}, Y.},
  \bibinfo{author}{{Bamford}, S.}, \bibinfo{author}{{Dickinson}, H.},
  \bibinfo{author}{{Fortson}, L.}, \bibinfo{author}{{Kruk}, S.},
  \bibinfo{author}{{Masters}, K.}, \bibinfo{author}{{Scarlata}, C.},
  \bibinfo{author}{{Simmons}, B.}, \bibinfo{author}{{Smethurst}, R.},
  \bibinfo{author}{{Wright}, D.}, \bibinfo{year}{2020}.
\newblock \bibinfo{title}{{Galaxy Zoo: probabilistic morphology through
  Bayesian CNNs and active learning}}.
\newblock \bibinfo{journal}{MNRAS} \bibinfo{volume}{491},
  \bibinfo{pages}{1554--1574}.
\newblock \DOIprefix\doi{10.1093/mnras/stz2816},
  \href{http://arxiv.org/abs/1905.07424}{ arXiv:1905.07424}.
\bibitem[{Wang and Deng(2018)}]{WD2018}
\bibinfo{author}{Wang, M.}, \bibinfo{author}{Deng, W.}, \bibinfo{year}{2018}.
\newblock \bibinfo{title}{Deep visual domain adaptation: A survey}.
\newblock \bibinfo{journal}{Neurocomputing} \bibinfo{volume}{312},
  \bibinfo{pages}{135--153}.
\newblock \DOIprefix\doi{10.1016/j.neucom.2018.05.083}.
\bibitem[{Wattenberg et~al.(2016)Wattenberg, Viégas and Johnson}]{WV2016}
\bibinfo{author}{Wattenberg, M.}, \bibinfo{author}{Viégas, F.},
  \bibinfo{author}{Johnson, I.}, \bibinfo{year}{2016}.
\newblock \bibinfo{title}{How to use t-sne effectively}.
\newblock \bibinfo{journal}{Distill} \URLprefix
  \url{http://distill.pub/2016/misread-tsne},
  \DOIprefix\doi{10.23915/distill.00002}.
\bibitem[{{Willett} et~al.(2017){Willett}, {Galloway}, {Bamford}, {Lintott},
  {Masters}, {Scarlata}, {Simmons}, {Beck}, {Cardamone}, {Cheung}
  et~al.}]{WG2017}
\bibinfo{author}{{Willett}, K.W.}, \bibinfo{author}{{Galloway}, M.A.},
  \bibinfo{author}{{Bamford}, S.P.}, \bibinfo{author}{{Lintott}, C.J.},
  \bibinfo{author}{{Masters}, K.L.}, \bibinfo{author}{{Scarlata}, C.},
  \bibinfo{author}{{Simmons}, B.D.}, \bibinfo{author}{{Beck}, M.},
  \bibinfo{author}{{Cardamone}, C.N.}, \bibinfo{author}{{Cheung}, E.}, et~al.,
  \bibinfo{year}{2017}.
\newblock \bibinfo{title}{{Galaxy Zoo: morphological classifications for 120
  000 galaxies in HST legacy imaging}}.
\newblock \bibinfo{journal}{MNRAS} \bibinfo{volume}{464},
  \bibinfo{pages}{4176--4203}.
\newblock \DOIprefix\doi{10.1093/mnras/stw2568},
  \href{http://arxiv.org/abs/1610.03068}{ arXiv:1610.03068}.
\bibitem[{{Willett} et~al.(2013){Willett}, {Lintott}, {Bamford}, {Masters},
  {Simmons}, {Casteels}, {Edmondson}, {Fortson}, {Kaviraj}, {Keel}
  et~al.}]{WL2013}
\bibinfo{author}{{Willett}, K.W.}, \bibinfo{author}{{Lintott}, C.J.},
  \bibinfo{author}{{Bamford}, S.P.}, \bibinfo{author}{{Masters}, K.L.},
  \bibinfo{author}{{Simmons}, B.D.}, \bibinfo{author}{{Casteels}, K.R.V.},
  \bibinfo{author}{{Edmondson}, E.M.}, \bibinfo{author}{{Fortson}, L.F.},
  \bibinfo{author}{{Kaviraj}, S.}, \bibinfo{author}{{Keel}, W.C.}, et~al.,
  \bibinfo{year}{2013}.
\newblock \bibinfo{title}{{Galaxy Zoo 2: detailed morphological classifications
  for 304 122 galaxies from the Sloan Digital Sky Survey}}.
\newblock \bibinfo{journal}{MNRAS} \bibinfo{volume}{435},
  \bibinfo{pages}{2835--2860}.
\newblock \DOIprefix\doi{10.1093/mnras/stt1458},
  \href{http://arxiv.org/abs/1308.3496}{ arXiv:1308.3496}.
\bibitem[{Wilson and Cook(2020)}]{GD2020}
\bibinfo{author}{Wilson, G.}, \bibinfo{author}{Cook, D.J.},
  \bibinfo{year}{2020}.
\newblock \bibinfo{title}{A survey of unsupervised deep domain adaptation}.
\newblock \bibinfo{journal}{ACM Transactions on Intelligent Systems and
  Technology} \bibinfo{volume}{11}.
\newblock \DOIprefix\doi{10.1145/3400066}.
\bibitem[{{Wu} et~al.(2018){Wu}, {Xiong}, {Yu} and {Lin}}]{WX2018}
\bibinfo{author}{{Wu}, Z.}, \bibinfo{author}{{Xiong}, Y.},
  \bibinfo{author}{{Yu}, S.}, \bibinfo{author}{{Lin}, D.},
  \bibinfo{year}{2018}.
\newblock \bibinfo{title}{{Unsupervised Feature Learning via Non-Parametric
  Instance-level Discrimination}}.
\newblock \bibinfo{journal}{arXiv e-prints} ,
  \bibinfo{pages}{arXiv:1805.01978}\DOIprefix\doi{10.48550/arXiv.1805.01978},
  \href{http://arxiv.org/abs/1805.01978}{ arXiv:1805.01978}.
\bibitem[{{Xu} et~al.(2021){Xu}, {Yang}, {Cao}, {Li}, {Mao} and
  {Chen}}]{XJ2021}
\bibinfo{author}{{Xu}, Y.}, \bibinfo{author}{{Yang}, J.},
  \bibinfo{author}{{Cao}, H.}, \bibinfo{author}{{Li}, Q.},
  \bibinfo{author}{{Mao}, K.}, \bibinfo{author}{{Chen}, Z.},
  \bibinfo{year}{2021}.
\newblock \bibinfo{title}{Partial video domain adaptation with partial
  adversarial temporal attentive network}.
\newblock \bibinfo{journal}{2021 IEEE/CVF International Conference on Computer
  Vision (ICCV)} , \bibinfo{pages}{9312--9321}.
\bibitem[{{York} et~al.(2000){York}, {Adelman}, {Anderson}, {Anderson},
  {Annis}, {Bahcall}, {Bakken}, {Barkhouser}, {Bastian}, {Berman}, et~al. and
  {SDSS Collaboration}}]{YA2000}
\bibinfo{author}{{York}, D.G.}, \bibinfo{author}{{Adelman}, J.},
  \bibinfo{author}{{Anderson}, John~E., J.}, \bibinfo{author}{{Anderson},
  S.F.}, \bibinfo{author}{{Annis}, J.}, \bibinfo{author}{{Bahcall}, N.A.},
  \bibinfo{author}{{Bakken}, J.A.}, \bibinfo{author}{{Barkhouser}, R.},
  \bibinfo{author}{{Bastian}, S.}, \bibinfo{author}{{Berman}, E.},
  \bibinfo{author}{et~al.}, \bibinfo{author}{{SDSS Collaboration}},
  \bibinfo{year}{2000}.
\newblock \bibinfo{title}{{The Sloan Digital Sky Survey: Technical Summary}}.
\newblock \bibinfo{journal}{Astron. J.} \bibinfo{volume}{120},
  \bibinfo{pages}{1579--1587}.
\newblock \DOIprefix\doi{10.1086/301513},
  \href{http://arxiv.org/abs/astro-ph/0006396}{ arXiv:astro-ph/0006396}.
\bibitem[{{You} et~al.(2019){You}, {Long}, {Cao}, {Wang} and {Jordan}}]{YL2019}
\bibinfo{author}{{You}, K.}, \bibinfo{author}{{Long}, M.},
  \bibinfo{author}{{Cao}, Z.}, \bibinfo{author}{{Wang}, J.},
  \bibinfo{author}{{Jordan}, M.I.}, \bibinfo{year}{2019}.
\newblock \bibinfo{title}{Universal domain adaptation}, in:
  \bibinfo{booktitle}{2019 IEEE/CVF Conference on Computer Vision and Pattern
  Recognition (CVPR)}, pp. \bibinfo{pages}{2715--2724}.
\newblock \DOIprefix\doi{10.1109/CVPR.2019.00283}.
\bibitem[{{Yuan} et~al.(2022){Yuan}, {He} and {Jiang}}]{YH2022}
\bibinfo{author}{{Yuan}, Y.}, \bibinfo{author}{{He}, X.},
  \bibinfo{author}{{Jiang}, Z.}, \bibinfo{year}{2022}.
\newblock \bibinfo{title}{{Adaptive open domain recognition by coarse-to-fine
  prototype-based network}}.
\newblock \bibinfo{journal}{Pattern Recognition} \bibinfo{volume}{128},
  \bibinfo{pages}{108657}.
\newblock \DOIprefix\doi{10.1016/j.patcog.2022.108657}.
\bibitem[{Zellinger et~al.(2019)Zellinger, Moser, Grubinger, Lughofer,
  Natschläger and Saminger-Platz}]{ZM2019}
\bibinfo{author}{Zellinger, W.}, \bibinfo{author}{Moser, B.A.},
  \bibinfo{author}{Grubinger, T.}, \bibinfo{author}{Lughofer, E.},
  \bibinfo{author}{Natschläger, T.}, \bibinfo{author}{Saminger-Platz, S.},
  \bibinfo{year}{2019}.
\newblock \bibinfo{title}{Robust unsupervised domain adaptation for neural
  networks via moment alignment}.
\newblock \bibinfo{journal}{Information Sciences} \bibinfo{volume}{483},
  \bibinfo{pages}{174--191}.
\newblock \DOIprefix\doi{10.1016/j.ins.2019.01.025}.
\bibitem[{{Zhang} et~al.(2018){Zhang}, {Ding}, {Li} and {Ogunbona}}]{ZD2018}
\bibinfo{author}{{Zhang}, J.}, \bibinfo{author}{{Ding}, Z.},
  \bibinfo{author}{{Li}, W.}, \bibinfo{author}{{Ogunbona}, P.},
  \bibinfo{year}{2018}.
\newblock \bibinfo{title}{Importance weighted adversarial nets for partial
  domain adaptation}.
\newblock \bibinfo{journal}{2018 IEEE/CVF Conference on Computer Vision and
  Pattern Recognition} , \bibinfo{pages}{8156--8164}.

\end{thebibliography}

\end{document}